\documentclass[aps,prc,onecolumn,superscriptaddress,nofootinbib,amsmath,amssymb,showkeys,10pt]{revtex4-2}

\usepackage{graphicx}
\usepackage{booktabs}
\usepackage{tabularx}
\usepackage{tikz}
\usetikzlibrary{shapes.geometric}
\usepackage{comment}
\usepackage{color}
\usepackage{float}
\usepackage[colorlinks=true,citecolor=blue,linkcolor=blue,urlcolor=blue]{hyperref}

\newcommand{\ket}[1]{\left|#1\right\rangle}
\newcommand{\bra}[1]{\left\langle#1\right|}

\newenvironment{linenomath}{}{}

\begin{document}

\title{Toward the Centennial of Wigner's Supermultiplet Symmetry: an \textit{Ab Initio} Perspective}

\author{Phong Dang}
\email{pdang5@lsu.edu}
\affiliation{Department of Physics and Astronomy, Louisiana State University, Baton Rouge, LA 70803, USA}

\author{Feng Pan}
\affiliation{Department of Physics, School of Science, Huzhou Normal University, Huzhou 313000, China}
\affiliation{Department of Physics, Liaoning Normal University, Dalian 116029, China}
\affiliation{Department of Physics and Astronomy, Louisiana State University, Baton Rouge, LA 70803, USA}

\begin{abstract}
As the centennial of Wigner's supermultiplet theory approaches in 2037, this review offers a
modern perspective through the recently advanced \textit{ab initio} Symmetry Adapted Model (SAM).
We begin with the symmetry's empirical foundations in two-nucleon data and electroweak processes,
and trace its emergence from the strong force through effective field theories, the QCD large-$N_c$
limit, the similarity renormalization group, and lattice calculations.
We then follow its historical arc, from early limited engagement,
through its revived interest, to the modern algebraic and first-principles resurgence.
Central to this renaissance are new nullspace-based algorithms for the U(3) and U(4) coupling
and recoupling coefficients, which remove a bottleneck spanning nearly nine decades and are
released as the open-source libraries \texttt{u3lib} and \texttt{u4lib}.
Building on them, the SAM is the first large-scale \textit{ab initio} shell model to encode both
Elliott's spatial U(3) and Wigner's spin-isospin U(4) symmetries, resolving individual
supermultiplets that share the same quadratic Casimir eigenvalue.
Early results show that the spin-isospin symmetry strongly dominates high-fidelity chiral
interactions and suppresses spin-isospin polarizability in light nuclei.
We close with potential directions that the SAM is poised to advance within the next decade.
\end{abstract}

\keywords{Wigner's supermultiplet symmetry, SU(4) spin-isospin symmetry, nuclear \textit{ab initio} calculations, symmetry-adapted methods, nuclear shell model, chiral effective field theory, coupling coefficients}

\maketitle

\section{Introduction}
\label{sec:intro}

The vast universe is governed by four fundamental forces, gravity that pulls astronomical systems together, electromagnetism that drives electricity and rotates magnets, the weak force that underpins radioactive processes, and the strong force that holds matter at extremely small scale. Among these forces, the strong force, as conspicuous from its name, is the most powerful one, not only because it has shaped our universe since the very beginning in quark-gluon plasma, but also because it possesses such tremendous energy that, once unleashed, can power all infrastructures that serve humankind.

To date, the best theory that describes the strong force is the theory of quantum chromodynamics (QCD) within the Standard Model (SM). According to this model, matter consists of elementary particles, called fermions and bosons. The former is split into two families, namely quarks and leptons. There are six types of quarks, two of which---up ($u$) and down ($d$) quarks---bind together via the strong force to form protons ($uud$) and neutrons ($udd$). These nucleons then interact via the residual of the strong force---often referred to as the strong nuclear force---that leaks out from their internal structure to bind into atomic nuclei. This force works in tandem with the electrostatic Coulomb interaction to determine how many protons and neutrons are allowed to form nuclei, while the weak interaction governs how these nuclei transform to one another.

Despite such intricacy of the nuclear force, recurring emergent phenomena have been observed across the chart of nuclides and continued to be probed on both the theoretical and experimental frontiers \cite{LongRangePlan2023,Brown2025JPG}. From the early days of nuclear physics and quantum mechanics, it was discovered by Eugene Paul Wigner---a giant physicist of the 20th century---that the nuclear force is mostly identical between protons and neutrons, regardless of their respective spins \cite{Wigner1937PR,Wigner1939PR}. This idea was inspired by Heisenberg's discovery of the isospin symmetry \cite{Heisenberg1932ZP} and is now commonly referred to as the supermultiplet symmetry, the spin-isospin exchange symmetry, the $\rm SU(4)$ or $\rm U(4)$ symmetry (which reflects the mathematical structure underpinning the symmetry). Wigner's dedication to advancing our understanding of foundational physics ultimately led him to the most prestigious science award in the world---the Nobel Prize---in 1963 ``for his contributions to the theory of the atomic nucleus and the elementary particles, particularly through the discovery and application of fundamental symmetry principles'' \cite{Wigner1963}. In modern days, Wigner's work has continued to influence the state of the art, transcending across many fields beyond nuclear and particle physics, be it from quantum chemistry to condensed matter science, and from atomic physics to advanced quantum computing.

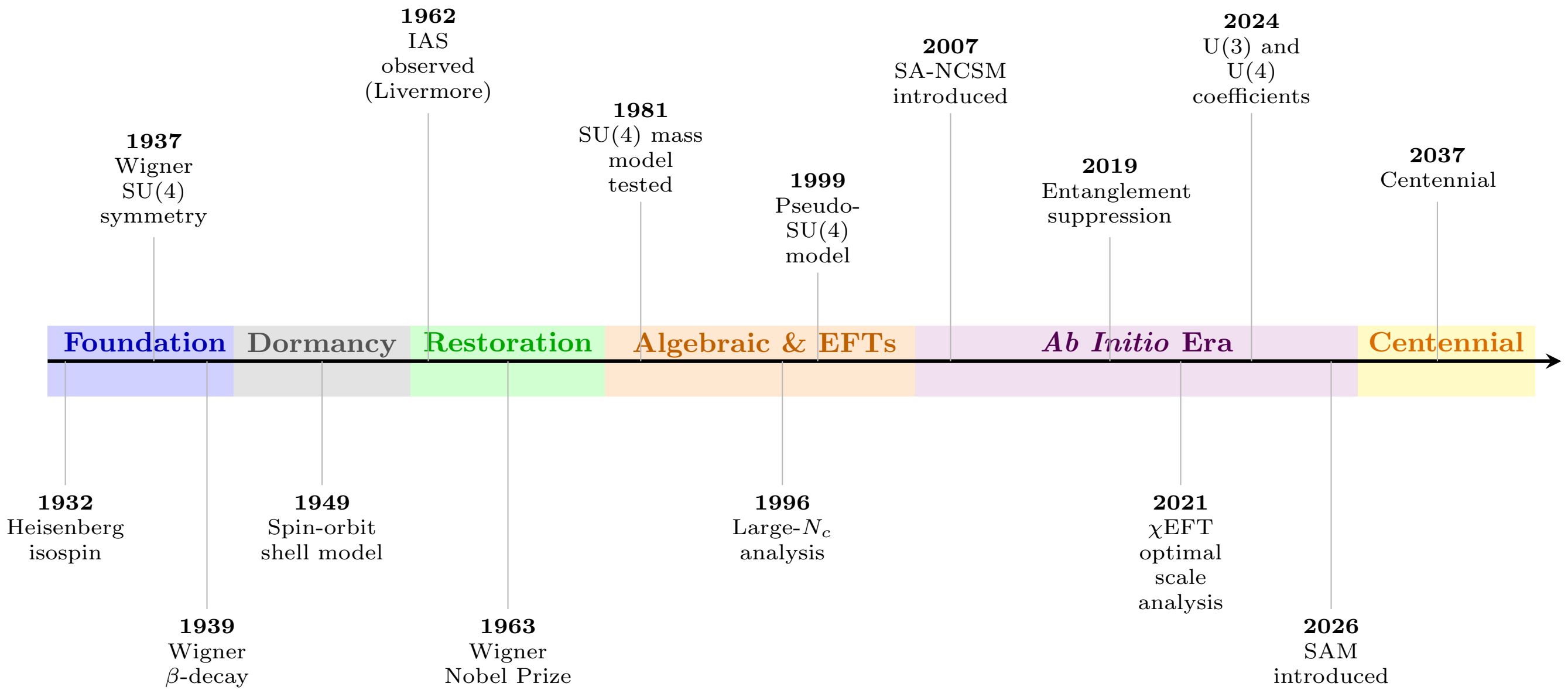
\begin{figure}[H]

\centering
\begin{tikzpicture}[
  font=\scriptsize,
  >=stealth,
  eA/.style={text width=1.55cm, align=center, anchor=south},
  eB/.style={text width=1.55cm, align=center, anchor=south},
  eC/.style={text width=1.55cm, align=center, anchor=north},
  eD/.style={text width=1.55cm, align=center, anchor=north},
  tick/.style={thin, gray!55},
]
\fill[blue!18]   (-0.3,-0.4) rectangle (1.8,  0.4);
\fill[gray!22]   (1.8, -0.4) rectangle (3.8,  0.4);
\fill[green!18]  (3.8, -0.4) rectangle (6.0,  0.4);
\fill[orange!18] (6.0, -0.4) rectangle (9.5,  0.4);
\fill[violet!12] (9.5, -0.4) rectangle (14.5, 0.4);
\fill[yellow!28] (14.5,-0.4) rectangle (16.5, 0.4);
\node[font=\bfseries,blue!70!black]    at (0.8,  0.22) {Foundation};
\node[font=\bfseries,gray!65!black]    at (2.8,   0.18) {Dormancy};
\node[font=\bfseries,green!65!black]   at (4.9,   0.22) {Restoration};
\node[font=\bfseries,orange!75!black]  at (7.8,  0.18) {Algebraic \& EFTs};
\node[font=\bfseries,violet!65!black]  at (12.0,  0.22) {\textit{Ab Initio} Era};
\node[font=\bfseries,orange!85!black]  at (15.5, 0.22) {Centennial};
\draw[very thick,->] (-0.3,0) -- (16.8,0);


\draw[tick] (0.9,0) -- (0.9,1.4);
\node[eA] at (0.9,1.4) {\textbf{1937}\\Wigner SU(4)\\symmetry};

\draw[tick] (4.0,0) -- (4.0,2.8);
\node[eB] at (4.0,2.8) {\textbf{1962}\\IAS observed\\(Livermore)};

\draw[tick] (6.4,0) -- (6.4,1.8);
\node[eA] at (6.4,1.8) {\textbf{1981}\\SU(4) mass\\model tested};

\draw[tick] (8.4,0) -- (8.4,1.0);
\node[eA] at (8.4,1.0) {\textbf{1999}\\Pseudo-SU(4)\\model};

\draw[tick] (9.9,0) -- (9.9,2.8);
\node[eB] at (9.9,2.8) {\textbf{2007}\\SA-NCSM\\introduced};

\draw[tick] (11.7,0) -- (11.7,1.4);
\node[eA] at (11.7,1.4) {\textbf{2019}\\Entanglement\\suppression};

\draw[tick] (13.3,0) -- (13.3,2.8);
\node[eB] at (13.3,2.8) {\textbf{2024}\\U(3) and U(4)\\coefficients};

\draw[tick] (15.4,0) -- (15.4,1.8);
\node[eA,text width=1.3cm] at (15.4,1.85) {\textbf{2037}\\Centennial};


\draw[tick] (-0.1,0) -- (-0.1,-1.4);
\node[eC] at (-0.1,-1.4) {\textbf{1932}\\Heisenberg\\isospin};

\draw[tick] (1.5,0) -- (1.5,-2.8);
\node[eD] at (1.5,-2.8) {\textbf{1939}\\Wigner\\$\beta$-decay};

\draw[tick] (2.8,0) -- (2.8,-1.4);
\node[eC] at (2.8,-1.4) {\textbf{1949}\\Spin-orbit\\shell model};

\draw[tick] (4.9,0) -- (4.9,-2.8);
\node[eD] at (4.9,-2.8) {\textbf{1963}\\Wigner\\Nobel Prize};

\draw[tick] (8.0,0) -- (8.0,-1.4);
\node[eC] at (8.0,-1.4) {\textbf{1996}\\Large-$N_c$\\analysis};


\draw[tick] (12.5,0) -- (12.5,-1.4);
\node[eC] at (12.5,-1.4) {\textbf{2021}\\$\chi$EFT optimal\\scale analysis};

\draw[tick] (14.2,0) -- (14.2,-2.8);
\node[eD] at (14.2,-2.8) {\textbf{2026}\\SAM\\introduced};

\end{tikzpicture}

\caption{Key milestones in the history of Wigner's supermultiplet symmetry, organized
by era. 
}
\label{fig:timeline}
\end{figure}

Over the nine decades since its proposal, interest in the supermultiplet symmetry has
experienced a remarkable trajectory with the key milestones in its history summarized in Figure~\ref{fig:timeline}. The 1940s and 1950s saw limited engagement with the
idea, largely because the triumph of the spin-orbit shell model of Mayer \cite{Mayer1948PR,Mayer1949PR}, Jensen, Haxel and Suess \cite{Haxel1949PR}---which explicitly relies on the spin-orbit
force that breaks SU(4)---shifted community attention toward a different paradigm.
Interest was revived dramatically in 1962 when the first observation of isobaric analog
states (IASs) in medium-mass nuclei at Lawrence Livermore National Laboratory in the U.S. \cite{Anderson1962PR} provided striking
evidence for a restoration of the symmetry in neutron-rich systems.
In the decades that followed, connections to QCD were illuminated through effective field theories (EFT) of the nuclear force in the limit of many colors (large $N_c$)  \cite{Kaplan1996PLB,Kaplan1997PRC} and large scattering length \cite{Mehen1999PRL}, chiral EFT ($\chi$EFT) \cite{Lee2021PRL}, similarity renormalization group (SRG) \cite{Cordon2008PRC,Timoteo2012PRC,Arriola2012PoS,Arriola2014PLB,Arriola2016Symmetry} and QCD lattice calculations at the
SU(3)-flavor symmetric point \cite{Beane2013PRD,Beane2013PRC}, and more recently
through entanglement suppression arguments \cite{Beane2019PRL}. Notably, the symmetry has been exploited in lattice simulations of nuclei and nuclear matter \cite{Lu2019PLB,Lu2020PRL,Shen2021EPJA,Ren2024PLB,Ren2025PRL,Lee2025ARNPS} to tame the sign problem of quantum Monte Carlo (QMC). Meanwhile, algebraic models continued to develop applications of the symmetry in heavy nuclei through the pseudo- and proxy-SU(4) schemes \cite{Isacker2023Symmetry,Kota2024PScr}. Most recently signatures of Wigner's symmetry were identified in no-core shell model (NCSM) wave functions through the quadratic Casimir invariant and its impact on $\beta$-decay has been probed \cite{Muli2026PRL}.
Yet a practical \textit{ab initio} framework that could leverage the symmetry directly in large-scale, realistic shell-model calculations had remained elusive.

This review is motivated by two converging circumstances, namely the approaching centennial of
Wigner's discovery in 2037 and the very recent introduction of the \textit{ab initio}
Symmetry Adapted Model (SAM) \cite{Dang2026}, the first large-scale nuclear structure
model that simultaneously encodes both the spatial U(3) (by Elliott) and the spin-isospin U(4) (by Wigner) symmetries.
The SAM is built upon enabling technologies developed over the past three years,
most notably new generic algorithms for computing coupling and recoupling coefficients
of the U(3) \cite{Dang2024EPJP1} and U(4) \cite{Dang2024EPJP2} groups, which had remained outstanding challenges for nearly nine decades and stalled the development of a truly comprehensive shell-model theory of Wigner's symmetry.
The SAM and these coefficient engines are developed as part of \texttt{DeepSANE} (\textbf{Deep}-learning and \textbf{S}ymmetry-adapted \textbf{A}b-initio \textbf{N}uclear modeling, an AI-ready \textbf{E}cosystem)\footnote{See the \texttt{DeepSANE} GitHub organization at \url{https://github.com/Deep-SANE}.}, an open-source ecosystem for symmetry-adapted \textit{ab initio} nuclear modeling that we are actively building.

The remainder of this review is organized as follows.
Section~\ref{sec:foundations} establishes the empirical and theoretical foundations of
Wigner's symmetry.
Section~\ref{sec:history} traces its historical arc across the nuclear chart.
Section~\ref{sec:abinitio} reviews the \textit{ab initio} symmetry-adapted approaches that set the stage for the SAM.
Section~\ref{sec:coefficients} describes the new coupling and recoupling coefficient algorithms as the key enabling technology.
Section~\ref{sec:SAM} presents the SAM and implications of its early results.
Section~\ref{sec:conclusions} summarizes the review and lays out a few directions that the SAM has the potential to address within the next decade.

\section{Foundations of Wigner's Supermultiplet Symmetry}
\label{sec:foundations}

\subsection{Empirical Evidence from the Two-Nucleon System}
\label{subsec:evidence}

The basic notion of the supermultiplet symmetry is that the nuclear force depends
mostly on spatial coordinates \cite{Wigner1937PR} and remains relatively invariant under transformations with respect to spin
and isospin. There is a wide range of examples that demonstrate the
existence of such approximate symmetry. In this subsection, we review some empirical
evidence of Wigner's symmetry from binding energy and scattering data of the two-nucleon
system at low energy (typically below a few hundred MeV of laboratory kinetic energy
and relative momentum), which inform the nature of the nuclear force. 

\begin{figure}[H]
\centering
\includegraphics[width=0.8\textwidth]{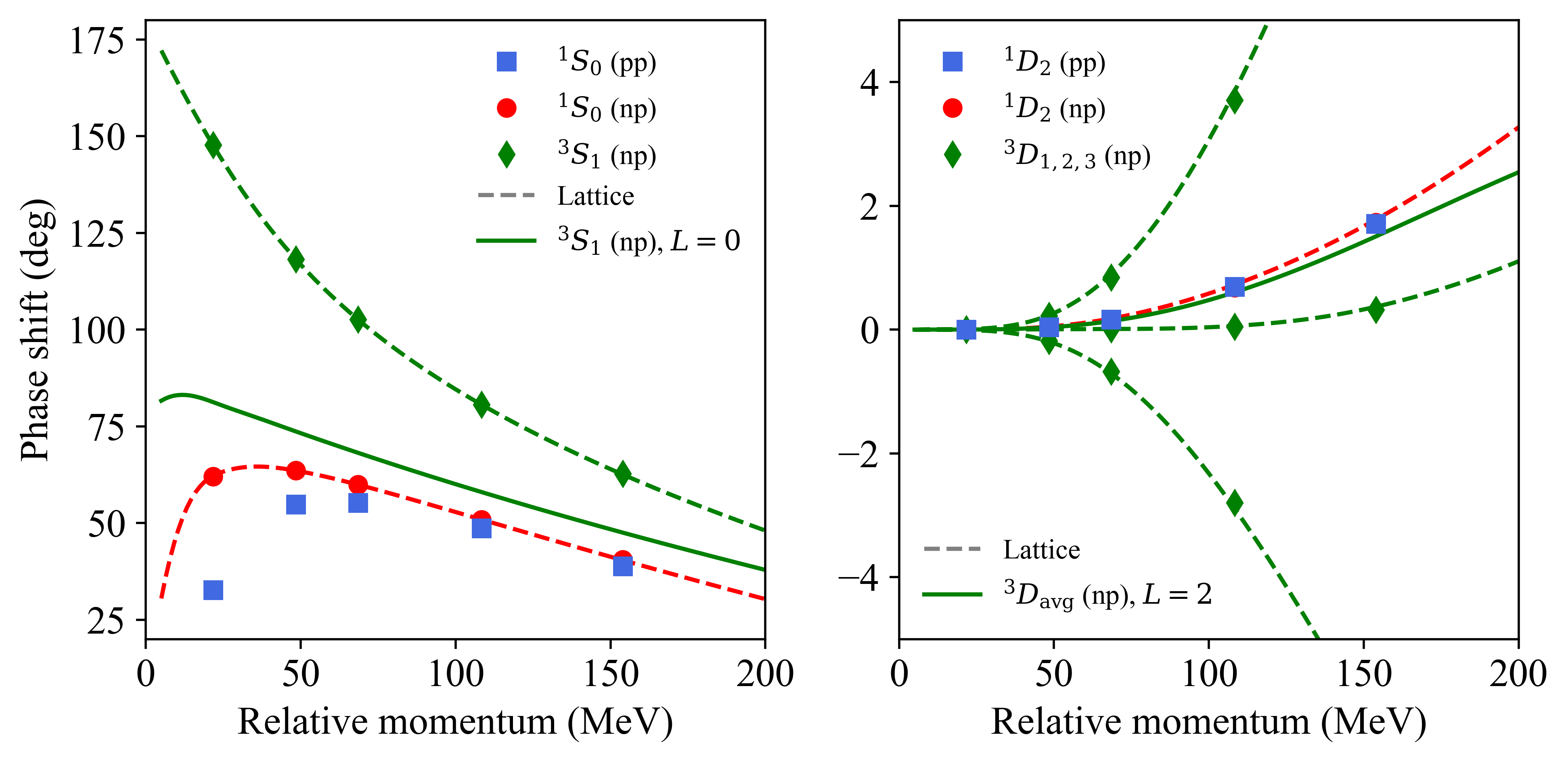}
\caption{Low-energy nucleon-nucleon scattering phase shifts as functions of relative momentum. The markers are data from Nijmegen partial wave analysis \cite{Stoks1993PRC}, whereas the dashed and solid lines are from np lattice calculation \cite{Lee2021PRL}. The left plot displays three $s$-wave channels; the dashed lines include spin-singlet $^1S_0$ (blue squares for pp and red circles for np) and spin-triplet channel $^3S_1$ (green diamonds for np); the solid green line is $^3S_1$ lattice phase shift restricted to $L=0$.  The right plot demonstrates $d$-wave phase shifts, wherein the spin-singlet channels $^1D_2$ for pp and np are drawn with blue squares and red circles, respectively, and the spin-triplet channels $^3D_J$ (with total angular momentum $J=1,2,3$) correspond to dashed lines with green diamonds. The solid green line represents the average of the three spin-triplet channels over all total angular momenta with orbital angular momentum restricted to $L=2$.}
\label{fig:phase_shift}
\end{figure}

\subsubsection{The nuclear force is almost spin symmetric}

In the low-energy regime, the scattering process between two nucleons is strongly
dominated by $s$-wave. For the system of a neutron and a proton (np), there are three spin-symmetric states
belonging to the spin-triplet channel $^3S_1$ and one spin-antisymmetric state associated
with the spin-singlet channel $^1S_0$. If the spin symmetry were an exact symmetry of the
nuclear force, these four states would be completely identical. Experimental
evidence shows that the nuclear force exhibits an approximate invariance with respect to spin.

First, let us consider the binding energy. The triplet channel forms the lightest composite nucleus,
the deuteron, with a small binding energy of $2.225$~MeV. Meanwhile, the singlet channel
is unbound, however, the threshold state is close to zero. This situation hints at the
presence of an approximate symmetry, since a tiny change in the nuclear force would make
the two channels become identical. Specifically, if the interaction gets slightly weaker,
the triplet states would be unbound; and if the interaction gains some strength, the
singlet state would be bound~\cite{Kaplan1996PLB,Lee2021PRL}.

Second, one may argue that the drastic difference between the scattering lengths
of the two channels, roughly $5.4$~fm for $^3S_1$ and $-20$~fm for $^1S_0$, indicates a severe breaking of the
symmetry. This, nonetheless, does not preclude the possibility of
an accidental symmetry of the nuclear force, since the scattering length can change dramatically even for small alteration of the nuclear potential 
in almost bound or almost unbound situation \cite{Kaplan1996PLB,Cordon2008PRC}, which is the case for the np system. 
Moreover, the fact that both the scattering lengths are 
unnaturally large as compared to the natural scale of the nuclear force---i.e., the Compton wavelength of
the pion, $\lambda_\pi = 1/M_\pi \approx 1.5$~fm \cite{Epelbaum2009RMP}---indicates that
the interaction is near the limit of Wigner's symmetry \cite{Chen2004PRL}. This is because
as the scattering length $a^{(s)}$, where $s$ represents the channel $^1S_0$ or $^3S_1$, goes to infinity,
the $s$-wave scattering amplitude~\cite{Mehen1999PRL}
\begin{linenomath}
\begin{equation}
  \mathcal{A}^{(s)} = -\frac{4\pi}{M}\frac{1}{1/a^{(s)} + ip}
  \label{eq:large-scattering}
\end{equation}
\end{linenomath}
becomes independent of the spin channel and depends solely on the nucleon momentum $p$,
reflecting an invariance with respect to spin.

Third, the signature of the spin symmetry can also be realized from np scattering phase
shifts. Figure~\ref{fig:phase_shift} shows phase shifts from Nijmegen
partial wave analysis~\cite{Stoks1993PRC} of high-precision nucleon-nucleon scattering and from nuclear lattice
calculations~\cite{Lee2021PRL}. As shown in the left plot, there is a relatively large
difference between the $^1S_0$ and $^3S_1$ np phase shifts, which occurs due to the presence
of the tensor force that mixes $^3S_1$ and $^3D_1$ channels, thereby masking the presence of the symmetry \cite{Lee2021PRL}. 
Once the effect of the tensor
force is removed by restricting lattice calculations to $L=0$ only~\cite{Lee2021PRL}, the singlet and triplet
phase shifts exhibit very similar shapes and, more importantly, are in the vicinity of one
another. Such a close proximity implies that a small modification in strength of the nuclear
force would make the two phase shifts analogous, hinting at the presence of the approximate
spin symmetry. Another compelling piece of evidence can also be found in $d$-wave scattering.
As it is illustrated in the right plot, the three
spin-triplet channels $^3D_{J=1,2,3}$ have very distinct phase shifts; specifically, two
of which are positive and the other is negative. However, one can average the phase shifts of the three channels
with lattice calculations being restricted to $L=2$~\cite{Lee2021PRL} (again, to remove the admixture generated by the
tensor force). It is remarkable that upon averaging over the three intrinsic spin orientations, the $d$-wave scattering phase shifts
are nearly identical between the spin-singlet and spin-triplet channels. 
A similar idea was demonstrated in Ref. \cite{Cordon2008PRC}, where the authors calculated the weighted mean phase shifts with the spin-orbit
and tensor force added through first-order perturbation. It was found that agreement between the singlet and triplet channels is valid up to roughly 250~MeV and 400~MeV of center-of-mass momentum for $d$- and $g$-waves, respectively, which also shows the long-distance nature of Wigner's symmetry, since sensitivity to short-distance effects decreases for higher partial waves \cite{Cordon2008PRC}. In addition, the authors provided evidence that the spin-isospin symmetry is strongly violated for odd-$L$ partial waves as suggested by the large-$N_c$ expansion.

\subsubsection{The nuclear force is almost isospin symmetric}

To examine the isospin symmetry of the nuclear interaction, one can consider a system of
two identical particles, either two protons (pp) or two neutrons (nn), which for $s$-wave
scattering, can only take the spin-singlet state, $^1S_0$, owing to antisymmetrization of
the wave function. Table~\ref{tab:pp_nn} shows the scattering lengths and effective range
parameters extracted from the pp and nn scattering; here, the former values have been
corrected for Coulomb repulsion to enable inference into the strong nuclear force only.
It is apparent that the gaps between pp and nn channels are very marginal, roughly 3\% for
the scattering lengths and 7\% for the effective ranges.

\begin{table}[H]
\centering
\caption{Scattering length and effective range of pp (Coulomb-corrected) and nn scattering, from~\cite{Krane1991}.}
\label{tab:pp_nn}
\begin{tabular}{ccc}  
\toprule
Channel & Scattering length (fm) & Effective range (fm) \\
\midrule
pp & $-17.1 \pm 0.2$ & $2.84 \pm 0.03$ \\
nn & $-16.6 \pm 0.5$ & $2.66 \pm 0.15$ \\
\bottomrule
\end{tabular} 
\end{table}


One can also realize the approximate isospin invariance of the nuclear force from scattering
phase shifts. In Figure~\ref{fig:phase_shift}, the two spin-singlet channels of pp and np
are nearly identical. In fact, for the $s$-waves, starting from around 50~MeV of relative
momentum, on average the difference between the phase shifts of the two channels is about
2.8 degrees, which is negligible compared to the scale of the phase shifts. Similarly, for
the $d$-waves, the deviations for relative momentum between 0 and 200~MeV are also very small.

\subsubsection{The nuclear force is almost spin-isospin symmetric}

In the above subsections, we have provided evidence for the spin symmetry and isospin
symmetry of the strong nuclear force separately. Combining all arguments regarding binding
energy, phase shifts, scattering lengths and effective ranges, one can arrive at the
conclusion that the low-energy nuclear interaction exhibits an approximate spin-isospin
exchange symmetry (Wigner's symmetry) \cite{Wigner1937PR}, which in its exact limit 
leads to the nuclear force treating the six $s$-wave spin-isospin states of the two-nucleon system on equal footing, see Figure~1 of Ref. \cite{Dang2026} for an illustration of this limit.

\subsection{Emergence of Wigner's Symmetry from the Strong Nuclear Force}
\label{subsec:emergence}

Since nuclear structure is an emergent phenomenon of the underlying quarks and
gluons, 
it is vital to understand how
Wigner’s spin-isospin symmetry emerges from such a complex structure.
To date, much work has been done to probe the emergence of the symmetry by means of the 
$\chi$EFT, along several complementary lines of investigation summarized below.

\subsubsection{Large-$N_c$ QCD and $\chi$EFT.}
Due to the non-perturbative nature of QCD at low energy, it is not possible to model the nuclear force starting from the quark and gluon degrees of freedom. Instead, the interactions between point-like protons and neutrons are modeled through the framework of EFTs.
In the late 90s, Kaplan, Savage, and Manohar~\cite{Kaplan1996PLB,Kaplan1997PRC} examined a
chiral Lagrangian of the two-nucleon system with two leading operators
\begin{linenomath}
\begin{equation}
  \mathcal{L} = -\frac{1}{2}C_S(N^\dagger N)^2 - \frac{1}{2}C_T(N^\dagger \boldsymbol{\sigma} N)^2,
\label{eq:Lagrangian}
\end{equation}
\end{linenomath}
where $N$ denotes isodoublet two-component spinors and $\boldsymbol{\sigma}$ represents
Pauli matrices. The first term is invariant under spin and isospin transformation, whereas
the second one breaks the symmetry. 
The authors found that the two-body
nuclear force is SU(4)-symmetric with corrections at the level of $C_T/C_S \approx 5\%$.
A more consequential finding surfaced when they analyzed diagrams representing baryon-baryon
interactions via exchange of quarks and gluons in the large-$N_c$ limit; while in reality, $N_c = 3$, this
technique simplifies structure of hadrons and by extension the analysis of their
properties. It was discovered that in this limit, interaction operators that do not preserve
the spin and isospin symmetries get suppressed as $1/N_c^2$, whereas spin- and
isospin-independent operators are combinatorially enhanced, thereby leading to the rise of
the exact supermultiplet symmetry as $N_c \rightarrow \infty$.

\subsubsection{Lattice QCD at the SU(3)-flavor symmetric point.}
Transitioning into the 21st century, with the availability of petascale supercomputers and
modern numerical techniques, state-of-the-art lattice QCD calculations of light nuclei were
performed by Beane \textit{et al.} at the limit of the SU(3)-flavor symmetry---where the up,
down and strange quarks are identical in their masses---using the physical strange-quark mass. 
It was found that there exist bound states for both spin-singlet and spin-triplet channels
with binding energies of 15.9~MeV for $^1S_0$ and 19.5~MeV for
$^3S_1$~\cite{Beane2013PRD} and scattering lengths of 2.33~fm for $^1S_0$ and 1.82~fm for
$^3S_1$~\cite{Beane2013PRC}. These results hint at the emergence of the supermultiplet
symmetry at the SU(3)-symmetric point of QCD.

\subsubsection{Entanglement suppression.}
More recently, Beane \textit{et al.}\ brought forward a fresh perspective on how the
symmetry arises from the strong nuclear force~\cite{Beane2019PRL}. To be
specific, they investigated the entanglement power of the S matrix in the spin space of the
np system using low-energy scattering data. Such a description allows for analysis of
entangled states that go beyond the product states discussed in
Section~\ref{subsec:evidence}. By identifying the vanishing entanglement power that
corresponds to the exact supermultiplet symmetry of the chiral Lagrangian, $C_T=0$ in
Eq.~(\ref{eq:Lagrangian}), the authors attributed the emergence of Wigner's symmetry to
entanglement suppression of the strong interaction in the infrared. 
It is worth noting that the underlying mechanism connecting these two phenomena is not yet known 
and was termed by the authors as a conjecture.

\subsubsection{Optimal resolution scale.} 
Mehen, Stewart, and Wise \cite{Mehen1999PRL} were the first to explore the scale at which Wigner's symmetry 
emerges from the effective two-body Lagrangian of the nucleon field. They found that in the limit of low momentum and large scattering length---i.e., $1/a^{(s)}  \ll p \ll \Lambda$, where $a^{(s)}$ is the same as in Eq. (\ref{eq:large-scattering}) and the scale $\Lambda$ was estimated to be about $400$~MeV---corrections to the symmetry are suppressed by ($1/a^{^1S_0} - 1/a^{^3S_1}$) and by powers of $p/\Lambda$ \cite{Mehen1999PRL}.
Lee \textit{et al.} recently showed that for the corrections of the symmetry to scale as $1/N_c^2$ as shown in Ref. \cite{Kaplan1996PLB},
the effective theory requires a so-called \textit{optimal} resolution scale of roughly $500$~MeV \cite{Lee2021PRL}. Interestingly, this momentum scale is also
where realistic interactions derived from the $\chi$EFT work most effectively,
hence, their finding strongly suggests that Wigner's symmetry and other emergent properties of the
strong interaction in the large-$N_c$ limit be appreciated in future nuclear structure calculations \cite{Lee2021PRL}.

\subsubsection{Renormalization of the nuclear force.} 
The emergence of Wigner's symmetry from renormalization scale of nuclear interactions 
has been investigated in a series of papers by Arriola and colleagues \cite{Cordon2008PRC,Timoteo2012PRC,Arriola2012PoS,Arriola2014PLB,Arriola2016Symmetry}.
In Refs. \cite{Timoteo2012PRC,Arriola2012PoS}, the authors examined the diagonal, $V(p,p)$, and fully off-diagonal, $V(p,0)$, matrix elements of nuclear interactions in the momentum space under evolution of the SRG flow. The main motivation underlying this idea is that the supermultiplet symmetry implies identical nuclear potentials in the $^1S_0$ and $^3S_1$ channels \cite{Cordon2008PRC}. By examining the Argonne AV18 potential \cite{Wiringa1995PRC} at various values of the SRG cutoff, the authors reported the realization of Wigner's symmetry at momentum scale of $\lambda \approx 3$~fm$^{-1} \approx 590$~MeV, which is in qualitative agreement with the resolution scales found previously in Refs. \cite{Mehen1999PRL,Lee2021PRL} and more importantly does not depend on details of the nuclear force \cite{Arriola2016Symmetry}.

\subsection{Wigner's Symmetry and Electroweak Processes}

So far in this section, we have considered only the lightest nuclear system composed of only two nucleons and their interaction.
A natural question arises: \textit{does the presence of Wigner's symmetry extend to more complex nuclei with more intricate structure?}
In fact, the most common example in support of the supermultiplet structure is
$\beta$-decay \cite{Wigner1939PR}, an isospin-flipping process that is generated by the coupling of nuclear structure 
to the weak force  and converts a proton
inside the nucleus into a neutron or vice versa. The reason as to why Wigner's symmetry is
vital for this process is grounded in the fact that the leading operators of the decay,
namely the Fermi operator of the vector current and the Gamow--Teller (GT) transition operator in the long-wavelength limit of the axial-vector current \cite{Baroni2018PRC,Muli2026PRL}, are generators of the SU(4) group, which
provides a mathematical framework for the symmetry and is presented later in Section~\ref{subsec:group_theory}. This implies
that $\beta$-decay takes place mostly within the same spin-isospin supermultiplet and is therefore largely protected by the overlap between the supermultiplet contents of the parent and daughter wave functions. Below, we show a few examples in Table \ref{tab:log-ft} from both theoretical and experimental perspectives that demonstrate the relevance of Wigner's symmetry.

\begin{table}[h!]
\caption{Examples that show the influence of Wigner's symmetry on $\beta$-decay. The first five rows subsume transitions with the parent and daughter nuclei sharing similar supermultiplet content, whereas the last three are decays between nuclei whose supermultiplet structures have small overlap. The $\log_{10}(ft)$ values are taken from Ref.  \cite{Tilley2002NPA} for the $\rm ^6He \rightarrow {^6Li}$ channel, Ref. \cite{Tilley2004NPA} for the $\rm ^8Li \rightarrow {^8Be}$ channel, Ref. \cite{Finlay2011PRL} for the $\rm ^{10}C \rightarrow {^{10}B^*}$ channel, and Ref. \cite{Kaplan1996PLB} for the $\rm ^{18}Ne \rightarrow {^{18}F} \rightarrow {^{18}O}$ channels.}
\label{tab:log-ft}
\begin{tabular}{ccc}  
\toprule
Decay & $(J^\pi,T) \rightarrow (J^\pi,T)$ & $\log_{10}(ft)$ \\
\midrule
$\rm ^6He \rightarrow {^6Li}$  & $(0^+,1) \rightarrow (1^+,0)$ & 2.910 \\
$\rm ^{10}C \rightarrow {^{10}B^*}$ (0.71 MeV)  & $(0^+,1) \rightarrow (1^+,0)$ & 3.483 \\
$\rm ^{18}F \rightarrow {^{18}O}$ & $(1^+,0) \rightarrow (0^+,1)$ & 3.554  \\
$\rm ^{18}Ne \rightarrow {^{18}F}$  & $(0^+,1) \rightarrow (1^+,0)$ & 3.096  \\
$\rm ^{18}Ne \rightarrow {^{18}F^*}$ (1.04 MeV) & $(0^+,1) \rightarrow (0^+,1)$ & 3.473  \\
\midrule
$\rm ^8Li \rightarrow {^8Be^*}$ (3.03~MeV) & $(2^+,1) \rightarrow (2^+,0)$ & $\ge$ 5.6  \\
$\rm ^{18}Ne \rightarrow {^{18}F^*}$ (1.08 MeV) & $(0^+,1) \rightarrow (0^-,0)$ & 7.012  \\
$\rm ^{18}Ne \rightarrow {^{18}F^*}$ (1.70 MeV) & $(0^+,1) \rightarrow (1^+,0)$ & 4.477  \\
\bottomrule
\end{tabular}  
\end{table}


Let us examine the first five examples in Table \ref{tab:log-ft}. State-of-the-art \textit{ab initio} calculations \cite{Muli2026PRL,Dang2026} and shell-model results \cite{Kaplan1996PLB} show that the ground states of $\rm ^6Li$, $\rm ^6He$, $\rm ^{10}C$, $\rm ^{18}Ne$, $\rm ^{18}F$, $\rm ^{18}O$ and excited states of $\rm ^{10}B^*$ at 0.71~MeV and $\rm ^{18}F^*$ at 1.04~MeV are strongly dominated by the 6-dimensional supermultiplet (corresponding to the $\rm U(4)$ representation $[1,1,0,0]$, see Section~\ref{subsec:group_theory}). These decay channels are those that occur mostly within the same supermultiplet. Hence, the fact that their $\log_{10}(ft)$ values are very similar supports the existence of Wigner's symmetry.

The excited state of $\rm ^{18}F^*$ at 1.70~MeV having the same $(J^\pi,T)$ with the ground state is also dominated by the same supermultiplet as the ground state. However, due to enhanced mixing with other supermultiplets in the wave function, decay from the ground state of $\rm ^{18}Ne$ to this excited state is about one order of magnitude smaller than that to the ground state of $\rm ^{18}F$. The first forbidden transition of $\rm ^{18}Ne$ to the excited state of $\rm ^{18}F^*$ at 1.08~MeV, which occurs outside the supermultiplet and involves a parity change, is much weaker, about four orders of magnitude smaller than that to the ground state \cite{Kaplan1996PLB}. Another example that has been recently shown in Ref. \cite{Muli2026PRL} is the decay from $\rm ^8Li$ ground state to the first excited state of $\rm ^8Be^*$. While the final state is predominantly a $\rm U(4)$ scalar (so is the ground state of $\rm ^8Be$ \cite{Dytrych2013PRL}), the initial state is not, simply because a $\rm U(4)$ scalar does not decompose to non-zero isospin \cite{Dang2024EPJP2}; therefore, the transition is roughly a hundred times weaker than the first five examples in Table \ref{tab:log-ft}.

Recent advance in the $\chi$EFT has enhanced the predictive capability of \textit{ab initio} calculations for $\beta$-decay rates with the introduction of the two-body axial current \cite{Gysbers2019NatP,King2020PRC,Glick-Magid2022PLB,King2025}. Remarkably, $\beta$-decay rates were predicted with high accuracy from tritium to tin by Gysbers \textit{et al.} \cite{Gysbers2019NatP} without the need for an empirical quenching factor \cite{Towner1987PR}. The first analysis of the impact that the two-body current has on the supermultiplet structure has been carried out recently by Muli \textit{et al.} for $\beta$-decays between $A=8$ systems \cite{Muli2026PRL}. It has been found that at the one-body level, the transitions are strongly suppressed due to small overlaps between the supermultiplet content of the wave functions (as explained in the previous paragraph); however, the two-body current can induce transition between different supermultiplets, thereby leading to a large two-body correction at the level of 30\% for these decays \cite{King2020PRC}. Moreover, the dominant components of the two-body current can be expressed in terms of two-body $\rm SU(4)$ generators; hence, these operators are diagonal under restriction to two-particle states, in which case their impacts are enhanced compared to the standard $\chi$EFT power counting \cite{Muli2026PRL}. A recent investigation of the $^3\mathrm{He}$ magnetic moment and the $^3\mathrm{H}$ GT matrix element within the pionless EFT found that Coulomb corrections are surprisingly smaller (by one to two orders of magnitude) than the expected EFT scaling \cite{Nguyen2026JPG}. Analyzing these corrections in the $\mathrm{SU}(4)$ basis revealed that the $^3\mathrm{He}$ magnetic moment correction is suppressed at the second order in symmetry breaking, bringing the revised power-counting estimate ($\sim 0.11\%$) into close agreement with the calculated $0.18\%$; for the $^3\mathrm{H}$ GT matrix element, however, $\mathrm{SU}(4)$ symmetry breaking only suppresses the correction to the first order ($\sim 0.92\%$), leaving its even smaller observed size ($0.075\%$) only partially explained.


The relationship between the supermultiplet symmetry and $\beta\beta$-decay was investigated earlier. 
In Ref. \cite{Vogel1986PRL}, Vogel and Zirnbauer examined the impact of the particle-particle
component of the spin-isospin polarization force on the two-neutrino $\beta\beta$-decay 
matrix element $M_{2\nu\beta\beta}$ between the $0^+$ ground states of two even-even nuclei
and found that the matrix element 
passes through zero as the particle-particle strength changes. To better understand this phenomenon, 
they investigated a schematic $\rm SO(8)$ Hamiltonian (Section \ref{subsec:apps} reviews the $\rm O(8)$ model)
and observed that this zero coincides with the 
exact $\rm SU(4)$ limit, where the particle-particle interaction strength becomes equal to the pairing one.
Moreover, the ground state of this Hamiltonian was shown to be the highest-weight state of a totally
symmetric $\rm SU(4)$ supermultiplet; thus the $\beta^+$ strength as well as $M_{2\nu\beta\beta}$ must vanish
in the $\rm SU(4)$ limit. Although Wigner's symmetry is broken in heavy nuclei,
the strong suppression of $M_{2\nu\beta\beta}$ demonstrated in this work elucidated why 
calculations of the matrix element are extremely challenging. 
In Ref. \cite{Vogel1988PLB}, Vogel, Ericson, and Vergados derived sum rules for the double GT 
and Fermi operators. Unlike $\beta$-decay, for which the GT and Fermi sum rules are exact and model-independent,
the double transition operators admit only approximate sum rules in general. 
These sum rules become exact in the limit of good isospin or good Wigner
$\rm SU(4)$ symmetry, and they remain a useful approximation for the Pauli-blocked nuclei with
$N-Z \gg 1$. The authors showed that the double GT operator feeds only the double-analogue 
and double GT states of the final even-even nucleus, which together exhaust the sum rule, 
while the double $\beta^+$ strength vanishes identically for $N > Z$. They further noted that 
amplitudes carrying the ``wrong'' isospin or $\rm SU(4)$ representation enter the sum rules only 
quadratically, so that a moderate symmetry breaking leaves them largely intact. 
In this way, the supermultiplet symmetry provides useful guidance for the
double charge-changing strength and sets the scale of its extreme suppression,
with the $2\nu$ mode exhausting only about $10^{-4}$ of the double GT sum rule.

\subsection{Group-Theoretical Framework}
\label{subsec:group_theory}

\subsubsection{The special unitary group SU(4)}

\begin{figure}[H]
\centering
\includegraphics[width=0.46\textwidth]{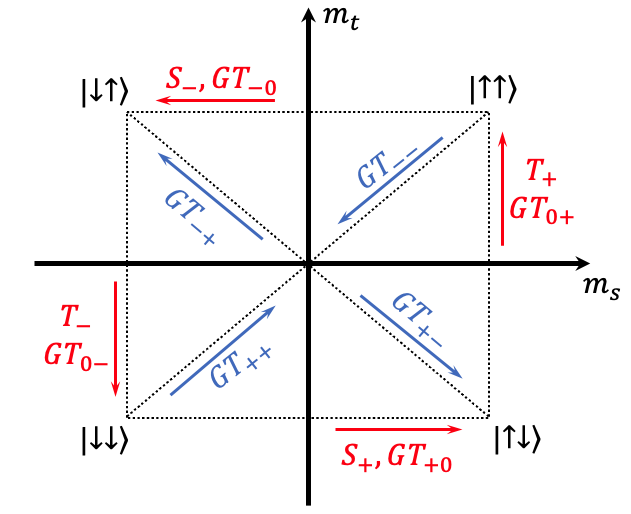}
\caption{Weight diagram of the fundamental representation of SU(4). The four spin-isospin single-particle states transform as an SU(4) quartet through
the 15 group generators.}
\label{fig:SU4_quartet}
\end{figure}

Let us temporarily set aside the spatial degrees of freedom of a nucleon (which will be put back later in Section~\ref{sec:SAM}), then there are
four spin-isospin states, $\ket{m_s m_t}$, that it can occupy, namely
$\ket{\uparrow\uparrow}$, $\ket{\uparrow\downarrow}$, $\ket{\downarrow\uparrow}$,
$\ket{\downarrow\downarrow}$. These states form the fundamental representation of the
special unitary group, SU(4), and transform as an SU(4) quartet via the action of 15 group
generators which are tensor components of the spin vector $\mathbf{S}_{m_s}$, the isospin vector
$\mathbf{T}_{m_t}$ and the double tensor Gamow--Teller $\mathbf{GT}_{m_s m_t}$.
Figure~\ref{fig:SU4_quartet} illustrates the action of these generators, which can be
classified into five groups:
\begin{itemize}
  \item raising operators: $S_+$, $GT_{+0}$, $T_+$, $GT_{0+}$, $GT_{++}$;
  \item lowering operators: $S_-$, $GT_{-0}$, $T_-$, $GT_{0-}$, $GT_{--}$;
  \item spin-raising isospin-lowering operator: $GT_{+-}$;
  \item spin-lowering isospin-raising operator: $GT_{-+}$;
  \item spin-isospin-preserving operators: $S_0$, $T_0$, $GT_{00}$.
\end{itemize}
The spin operators $\{S_+, S_0, S_-\}$ and isospin operators $\{T_+, T_0, T_-\}$ form
two subgroups $\text{SU}_S(2)$ and $\text{SU}_T(2)$, leading to the group-subgroup chain
$\text{SU}(4) \supset \text{SU}_S(2) \otimes \text{SU}_T(2)$.

\subsubsection{Classification of nuclear states}

Going beyond a single particle, the management of the symmetry can be easily achieved by
adding the number operator $A$---also called the baryon number \cite{Mehen1999PRL}---that counts the total number of nucleons in the system.
The number operator together with the 15 generators listed above form the unitary group,
U(4), and quantum states of a nucleus can be labeled as
\begin{linenomath}
\begin{equation}
  \ket{[n_{14} n_{24} n_{34} n_{44}] \; \zeta \; S \; M_S \; T \; M_T},
\label{eq:su4su2su2}
\end{equation}
\end{linenomath}
The first four non-negative integers ($n_{14} \ge n_{24} \ge n_{34} \ge n_{44} \ge 0$) form a U(4) irreducible representation (irrep) and represent the
maximum numbers of nucleons that can be distributed into each single-particle spin-isospin
state \cite{Dang2024EPJP2}. The last four quantum numbers simply denote the collective spin and isospin of the
entire nucleus and their projections. Since within a $\rm U(4)$ irrep the same pair $(S,T)$ can occur multiple times, one must introduce additional labels to keep track of this degeneracy. Former work by Draayer \cite{Draayer1970JMP} introduced two multiplicity labels, one for spin and one for isospin, through a similar projection procedure that was used for constructing orbital angular momentum label for Elliott's $\rm SU(3)$ symmetry \cite{Draayer1968NPA}. Another work done by Hecht and Pang \cite{Hecht1969JMP} sidestepped this issue by considering only $\rm U(4)$ irreps that have no degeneracy with respect to $S$ and $T$. Here, we adopt a single multiplicity label, $\zeta$, which was first advanced by Racah \cite{Racah1949RMP,Quintero2026CPC} and has recently been used to construct $\rm U(4) \supset SU_S(2) \otimes SU_T(2)$ basis states in Refs. \cite{Pan2023EPJP,Pan2023NPA,Pan2024CPC,Dang2024EPJP2}.

It is important to note that while some authors adopt this labeling
scheme using $\rm U(4)$ irrep \cite{Pan2023EPJP,Pan2023NPA,Pan2024CPC,Dang2023PRC,Dang2024EPJP2}, one more often encounters in
the literature the symmetry being represented by SU(4) irreps with three quantum numbers
$(PP'P'')$, which is related to U(4) irreps via a simple conversion
\begin{itemize}
\item $P = (n_{14} + n_{24} - n_{34} - n_{44})/2, \quad P' = (n_{14} - n_{24} + n_{34} - n_{44})/2 \text{ and } P'' = (n_{14} - n_{24} - n_{34} + n_{44})/2$ according to the original convention by Wigner \cite{Hecht1969JMP};
\item $P = n_{14} - n_{24}, \quad P' = n_{24} - n_{34} \quad \text{and} \quad P'' = n_{34} - n_{44}$ according to Draayer's convention \cite{Draayer1970JMP}.
\end{itemize}
While these two schemes of irrep labeling preserve the spin-isospin content of the nuclear system, they remove the
information on the nucleon number, $A = \sum_i n_{i4}$. It is the choice of developers to choose whichever scheme is most convenient.

The Hermitian conjugate of a tensor that transforms according to representation of the $\rm U(4)$ group is obtained as follows. For a $\rm U(4)$ tensor of rank $[n_{14} n_{24} n_{34} n_{44}]$, its conjugate tensor has rank $[n_{14}-n_{44} \, n_{14}-n_{34} \, n_{14}-n_{24} \, 0]$. Using the above definitions of $\rm SU(4)$ irrep, the conjugation of $\rm SU(4)$ tensor obeys the following rules: $(P P' P'') \rightarrow (P P' {-P''})$ for Wigner's convention, and $(P P' P'') \rightarrow (P'' P' P)$ for Draayer's convention.

\subsubsection{Hamiltonian with Wigner's symmetry}

By definition, a nuclear Hamiltonian that is symmetric with respect to spin-isospin transformations must commute with all generators of the $\rm SU(4)$ group, i.e.,
\begin{equation}
[ H, \mathbf{S} ] = [H, \mathbf{T} ] = [H, \mathbf{GT} ] = 0,
\end{equation}
where $\mathbf{S}$, $\mathbf{T}$ and $\mathbf{GT}$ are summed over the nucleons of a given nucleus. A notable example of such a Hamiltonian is one of Elliott's $\rm SU(3)$ type \cite{Elliott1958PRSA,Elliott1958PRSA-II,Elliott1963PRSA-III,McCoy2020PRL} that is made of products and powers of the $\rm SU(3)$ group generators, known as the Integrity Basis Operators \cite{Draayer1985NPA,Gilmore1985JMP,Leschber1986JPG} (which also include the Casimir invariants). This kind of Hamiltonian is completely spatial and thus invariant under rotations in the spin-isospin space. A less known set of spin and isospin scalars was also developed in Ref. \cite{Quesne1976JMP}, which can also be included in an $\rm SU(4)$-symmetric nuclear Hamiltonian.  

\subsubsection{Casimir invariants}

For any (special) unitary group, Casimir invariants are operators that are made of products of group generators and commute with all of them. Their eigenvalues go as polynomials of the irrep labels and are independent of subgroup quantum numbers. For a ${\rm U}(N)$ [${\rm SU}(N)$] group, there are $N$ [$N-1$] Casimir operators. In the simplest case of angular momentum algebra $\rm SU(2)$, the operator $J^2$ is of degree 2 and the only Casimir operator of this special unitary group. For the $\rm U(4)$ group, there are four Casimir operators; the first-order one is the number operator $\langle \mathfrak{C}_1 \rangle = A = n_{14} + n_{24} + n_{34} + n_{44}$; the quadratic, cubic, quartic operators are defined as follows
\begin{align}
\label{eq:U4_Casimir}
    \mathfrak{C}_2 = E_{ij} E_{ji} \text{, } \mathfrak{C}_3 = E_{ij} E_{jk} E_{ki} \text{ and } \mathfrak{C}_4 = E_{ij} E_{jk} E_{kl} E_{li},
\end{align}
where $E_{ij}$ ($1 \le i,j \le 4$) are the sixteen canonical generators in the group-subgroup chain $\rm U(4) \supset U(3) \supset U(2) \supset U(1)$ \cite{Dang2024EPJP2,Alex2011JMP} and summation over double indices is implied. Using techniques in Appendix B of Ref. \cite{Kota2020} and matrix elements given in Ref. \cite{Dang2024EPJP2}, we arrive at the following eigenvalues of the Casimirs for a $\rm U(4)$ irrep
\begin{align}
\langle \mathfrak{C}_2 \rangle = & n_{14}(n_{14}+3) + n_{24}(n_{24}+1) + n_{34}(n_{34}-1) + n_{44}(n_{44}-3),\\
\langle \mathfrak{C}_3 \rangle = & n_{14}^3 + n_{24}^3 + n_{34}^3 + n_{44}^3 + 6 n_{14}^2 + 3 n_{24}^2 - 3 n_{44}^2 \nonumber\\
						&  - n_{14}(n_{24}+n_{34}+n_{44}) - n_{24}(n_{34}+n_{44}) - n_{34} n_{44} \nonumber\\
						& + 9 n_{14} - n_{24} - 5 n_{34} - 3 n_{44}, \\
\langle \mathfrak{C}_4 \rangle = & n_{14}^4 + n_{24}^4 + n_{34}^4 + n_{44}^4 + 9n_{14}^3 + 5n_{24}^3 + n_{34}^3 - 3n_{44}^3 \nonumber\\
						 & - n_{14}^2 (n_{24} + n_{34} + n_{44} - 27) - n_{24}^2 (n_{14} + n_{34} + n_{44} - 5) \nonumber\\
						 & - n_{34}^2 (n_{14} + n_{24} + n_{44} + 5) - n_{44}^2 (n_{14} + n_{24} + n_{34} +3) \nonumber\\
						 & - n_{14}(8 n_{24} + 6 n_{34} + 4 n_{44} - 27) - n_{24}(4n_{34} + 2n_{44} + 11) - 13n_{34} - 3n_{44}.
\end{align}

Explicit formulas for the Casimir invariants of the $\rm SU(4)$ group were derived for Wigner's convention of the $\rm SU(4)$ labels by Burdet, Maguin and Partensky \cite{Burdet1968NuovoCim}, and by Danos and Gillet \cite{Danos1972ZP}. In Ref. \cite{Dang2026_2}, we derive the eigenvalues of these operators for Draayer's convention with a slightly different normalization. The eigenvalues of the $\rm SU(4)$ Casimirs are straightforward by a simple substitution from $\rm U(4)$ to $\rm SU(4)$ quantum numbers \cite{Kota2020}. We also impose the following conjugation relations for the Casimirs
\begin{align}
    \mathcal{C}_2(PP'P'') = \mathcal{C}_2(P''P'P) \text{, } \mathcal{C}_3(PP'P'') = -\mathcal{C}_3(P''P'P) \text{ and } \mathcal{C}_4(PP'P'') = \mathcal{C}_4(P''P'P)
\end{align} 
by invoking a linear mapping that preserves their algebraic properties
\begin{equation}
    \mathcal{C}_2 = 2 \mathfrak{C}_2 \text{, } \mathcal{C}_3 = 4(\mathfrak{C}_3 - \mathcal{C}_2) \text{ and } \mathcal{C}_4 = \mathfrak{C}_4 - \mathcal{C}_3.
\end{equation}
The final expressions for the $\rm SU(4)$ Casimir eigenvalues are
\begin{align}
    \langle \mathcal{C}_2 \rangle =& \frac{3}{2}(P^2+P''^2) + 2P'^2 + 2(P+P'') P' + P P'' + 6(P + P'') + 8P', \\
    \langle \mathcal{C}_3 \rangle =& \frac{3}{2} (P - P'')[P^2 + P''^2 + 2(P P' + P' P'' + P P'') + 2(3P + 2P' + 3P'' + 4)], \\
    \langle \mathcal{C}_4 \rangle =& \frac{1}{64} [21(P^4 + P''^4) + 16 P'^4 + 4P^3 (14 P' + 7 P'' + 42) + 4 P''^3 (14 P' + 7 P + 42) \nonumber \\
                   & + 32 P'^3 (4 + P + P'') + 72 P'^2 (P^2 + P''^2) + 30 P^2 P''^2 + 24 P P' P'' (3 P + 2 P' + 3 P'') \nonumber \\
                   & + (288 P'^2 + 896 P' + 960) (P + P'') + 24 P^2 (16 P' + 9 P'') + 24 P''^2 (16 P' + 9 P) \nonumber \\
                   & + 384 P P' P'' + 576 (P^2 + P''^2) + 512 P'^2 + 640 P P'' + 1024 P' ].
\end{align}

\section{Historical Overview of Wigner's Symmetry across the Nuclear Chart}
\label{sec:history}

\subsection{Early Skepticism and the Shell Model Era}

The initial reception of Wigner's supermultiplet theory was relatively cautious.
The 1940s and 1950s were dominated by the development of the nuclear shell model with
strong spin-orbit coupling~\cite{Mayer1948PR,Mayer1949PR,Haxel1949PR}, which earned
Mayer and Jensen a share of the 1963 Nobel Prize alongside Wigner.
The spin-orbit force explicitly breaks the $\rm SU(4)$ symmetry, and its vital role in reproducing
the observed nuclear magic numbers implies that the supermultiplet scheme would be badly broken
for nuclei beyond the $p$-shell.

\subsection{Restoration Through Isobaric Analog States}

The landscape changed dramatically in 1962 when Anderson, Wong, and McClure at Lawrence Livermore
National Laboratory reported the first observation of IASs in medium-mass nuclei~\cite{Anderson1962PR}.
IASs are states in neighboring isobars belonging to the same isospin multiplet and lying
close in energy, signaling a persistence of isospin (and by extension supermultiplet)
symmetry despite the spin-orbit force thought to be important in heavy isotopes.

One year later, Franzini and Radicati~\cite{Franzini1963PL} proposed a test of the symmetry using an
energy ratio
\begin{equation}
\label{eq:su4ratio}
 R(T_z) = \frac{E(A,T_z) - E(A,T_z-2)}{E(A,T_z-1) - E(A,T_z-2)},
\end{equation}
which was derived from a nuclear mass formula built on the quadratic Casimir invariant of the $\rm SU(4)$ group:
\begin{equation}
\label{eq:su4mass}
  E(A,T_z) = a(A) + b(A) \mathcal{C}_2,
\end{equation}
where the coefficients $a$ and $b$ are functions of the mass $A$.
If the supermultiplet model holds, this ratio should be independent of $A$. They found that for a broad range of
neutron-rich medium and heavy nuclei ($N-Z \gg 1$), the experimental values of this ratio agreed with the
theoretical predictions, providing the first confirmation of a restoration of Wigner's symmetry toward the neutron drip line.
Interestingly, Wigner's Nobel Prize followed the dramatic Livermore observation by just one year, 
though the prize recognized the full breadth of his symmetry contributions to physics.

\subsection{Applications and Tests of Algebraic Models 
}
\label{subsec:apps}

For the remainder of the 20th century, applications of the supermultiplet model grew
gradually. The initial nuclear mass formula based on the quadratic $\rm SU(4)$ Casimir in Eq. (\ref{eq:su4mass}) was refined 
with Coulomb and shell corrections \cite{Cauvin1981NPA}, higher-order Casimir terms \cite{Burdet1968NuovoCim,Gaponov1982NPA},
and Weizs\"acker-type contributions \cite{Isacker1997FoP}. In each case the $\rm SU(4)$
formula slightly outperformed the Weizs\"acker formula in root-mean-squared error. 
Around the same time, the ratio in Eq. (\ref{eq:su4ratio}) was shown by Chakraborty, Kota and Parikh to be insensitive to the breaking of the symmetry, and thus not an appropriate test \cite{Chakraborty1980PRL}. 
Later in the mid-90s, Van Isacker, Warner and Brenner proposed a more sensitive test with the double binding energy difference \cite{Isacker1995PRL}
\begin{equation}
\delta V_{np} (N,Z) = \frac{1}{4} \{ [ E(N,Z) - E(N-2,Z) ] - [ E(N,Z-2) - E(N-2, Z-2) ] \},
\end{equation}
which was known to be anomalously large for even-even self-conjugate nuclei. The substitution with binding energy in Eq. (\ref{eq:su4mass}) yielded a very intuitive explanation for this phenomenon and further supported the existence of Wigner's symmetry. Furthermore, the authors observed that for even-even self-conjugate nuclei in the $sd$-shell, as the mass increases the double binding energy difference first decreases and then increases back toward the end of the shell, which also implies a restoration of the supermultiplet symmetry \cite{Isacker1995PRL}. A few years later, the double binding energy difference was investigated for an infinite nuclear matter mass model \cite{Nayak2001PRC}.

More stringent tests of the symmetry were carried out via shell model calculations. 
Specifically, spectral distribution methods were used to confirm significant SU(4) breaking
in the $sd$-shell \cite{Hecht1974NPA,Haq1974NPA}. Notably, Vogel and Ormand \cite{Vogel1993PRC}
quantified the goodness of the symmetry by computing the overlap between shell-model wave functions 
of several $sd$- and $fp$-shell nuclei and the eigenstates of the quadratic $\rm SU(4)$ Casimir---a high
overlap indicates that Wigner's symmetry dominates the wave functions and a small overlap signals symmetry breaking 
(more comment on this approach will be presented in Section~\ref{sec:SAM} with the introduction of the SAM). 
In spite of the presence of symmetry breaking for most cases, the authors found a restoration of the supermultiplet structure
toward the neutron dripline along the calcium isotopic chain. 

Pairing correlations provide another setting in which Wigner's symmetry connects
nuclear structure to an algebraic description. The spin-isospin framework
accommodates both isovector and isoscalar pairing, allowing their competition to
be examined within a common space. This connection underlies the O(8) pairing
model and the interacting-boson-model (IBM) extension IBM-4, which includes
both types of pairs~\cite{Flowers1964FP18,Elliot1981FP26}. Beyond pairing, pseudo-
and proxy-SU(4) schemes extend the use of the supermultiplet framework to heavier
nuclei~\cite{Isacker1999PRL,Kota2024PScr}. Together, these developments illustrate
how Wigner's symmetry has continued to guide nuclear modeling even where it is
substantially broken.


The algebraic structure connecting these models can be formulated in the $LST$-coupling scheme, where orbital angular momentum, spin, and isospin are treated explicitly. Let $\{a^{\dag}_{\omega,\sigma,\tau},\,a_{\omega,\sigma,\tau}\}$ be a set of valence nucleon creation and annihilation operators in the spherical shell model, where $\omega=1,\,2,\,\cdots,\,\Omega\equiv\sum_{l}(2l+1)$ is the spatial index, and $\sigma$ and $\tau$, each taking a value of $1/2$ or $-1/2$, denote the third components of spin and isospin, respectively. The total number of many-particle product states ${\cal N}_{t}$ provided by $\{\Pi^{\dag}_{\phi_{k}}\vert0\rangle=\prod_{i=1}^{k}a^{\dag}_{\omega_{i},\sigma_{i},\tau_{i}}\vert0\rangle\}$, where $\vert0\rangle$ is the valence nucleon vacuum state and $\phi_{k}\equiv\{\omega_{1},\sigma_{1},\tau_{1},\cdots,\omega_{k},\sigma_{k},\tau_{k} \}$ denotes the full set of sub-indices involved, is given by
\begin{equation}
{\cal N}_{t}=\sum_{k=0}^{4\Omega}{(4\Omega)!\over{k!(4\Omega-k)!}}=2^{4\Omega}.
\end{equation}
This result is a direct consequence of the Pauli exclusion principle, which limits the maximal number of creation operators in any nonzero many-particle product state to
 $4\Omega$. 
The set of operators $\{Q_{\phi_{k},\phi'_{k'}}=\Pi^{\dag}_{\phi_{k}} \Pi_{\phi'_{k'}},\,0\leq k,k'\leq 4\Omega\}$ generates the unitary group U$(2^{4\Omega})$~\cite{mos,judd}. The set of many-particle product (Fock) states $\{ \Pi^{\dag}_{\phi_{1}}\vert0\rangle, \cdots, \Pi^{\dag}_{\phi_{4\Omega}}\vert0\rangle\}$ spans a complex linear space carrying the fundamental irreducible representation $[1,0,\cdots,0]$ of U$(2^{4\Omega})$. A subset $\{\Pi^{\dag}_{\phi_{k}}\,,\,\Pi_{\phi_{k}}\}$ with $k=1,2$ and $H_{\phi\phi'}=\Pi^{\dag}_{\phi_{1}}\Pi_{\phi'_{1}}$ generates the O$(8\Omega+1)$ group. Consequently, one has the embedding U$(2^{4\Omega})\supset$\,O$(8\Omega+1)$ with the branching rule $[1,0,\cdots,0]\downarrow({1\over{2}},\cdots,{1\over{2}})$, where $({1\over{2}},\cdots,{1\over{2}})$ with $4\Omega$ components equal to ${1\over{2}}$ denotes the fundamental spinor representation of O$(8\Omega+1)$.

The largest nontrivial subgroup of O$(8\Omega+1)$ is O$(8\Omega)$, generated by $\{\Pi^{\dag}_{\phi_{2}},\Pi_{\phi_{2}}, H_{\phi \phi' }\}$, with the branching rule:

\begin{equation}
\begin{array}{l}
{\rm O}(8\Omega+1) \downarrow~~{\rm O}(8\Omega)\cr
({1\over{2}},\cdots,{1\over{2}})\downarrow({1\over{2}},\cdots,{1\over{2}},{1\over{2}})\oplus
({1\over{2}},\cdots,{1\over{2}},-{1\over{2}}),
\end{array}
\end{equation}
where the representation $({1\over{2}},\cdots,{1\over{2}},{1\over{2}})$ of O$(8\Omega)$ is spanned by $\{ \Pi^{\dag}_{\phi_{k}}\vert0\rangle\}$ with $k$ even, while $({1\over{2}},\cdots,{1\over{2}},-{1\over{2}})$ is spanned by $\{ \Pi^{\dag}_{\phi_{k}}\vert0\rangle\}$ with $k$ odd. Two important reduction chains for O$(8\Omega)$ are O$(8\Omega)\supset$U$(4\Omega)$ and O$(8\Omega)\supset$O(8)$\otimes$O$(\Omega)$:

\begin{equation}\label{redu}
\begin{array}{l}
 ~~~~~~~~~_\nearrow \left[{\rm O}(8)\supset {\rm U}(4)
 \supset {\rm SU}_{S}(2)\otimes {\rm SU}_{T}(2)\right]
 \otimes {\rm O}(\Omega) \cr
{\rm O}(8\Omega)\cr
 ~~~~~~~~~^\searrow {\rm U}(4\Omega)\supset[{\rm U}(4)\supset {\rm SU}_{S}(2)\otimes {\rm SU}_{T}(2)]\otimes[{\rm U}(\Omega)\supset {\rm O}(\Omega) ].
\end{array}
\end{equation}

A partial list of the associated branching rules and other relevant group chains not shown above may be found in \cite{pang,Kota2006NPA,rowe06,kota16,kota23}. In the upper ${\rm O}(8)\otimes {\rm O}(\Omega)$ basis shown in (\ref{redu}), the reduction ${\rm O}(\Omega) \supset {\rm SO}_{L}(3)$ may be employed to classify basis vectors when $\Omega \geq3 $, where $L$ denotes the orbital angular momentum. In the lower ${\rm U}(4) \otimes {\rm U}(\Omega)$ basis shown in (\ref{redu}), one may adopt either 
${\rm U}(\Omega) \supset {\rm O}(\Omega) \supset {\rm SO}_{L}(3)$
or 
${\rm U}(\Omega)\supset {\rm SU}(3)\supset {\rm SO}_{L}(3)$ 
to classify the basis vectors when $\Omega \geq 3$ \cite{Elliott1958PRSA,Elliott1958PRSA-II,Elliott1963PRSA-III}. In either case, the reduction $\rm U(4) \supset SU_{S}(2) \otimes SU_{T}(2)$ is involved in both group chains shown in (\ref{redu}), making this framework particularly useful for nuclear shell-model calculations in the $LST$-coupling scheme.

It is well established that pairing interactions between like-nucleon pairs (i.e., nn and pp) play a crucial role across the entire nuclear chart, governing phenomena such as the pairing gap, odd--even mass staggering, and the structure of low-lying collective states. However, when the numbers of neutrons and protons are approximately equal ($N \approx Z$), an additional degree of freedom emerges: np pairing becomes energetically competitive and can significantly influence the nuclear dynamics in $N \approx Z$ nuclei. 
The O(8) model \cite{Flowers1964FP18} was introduced to describe isovector and isoscalar pairing interactions restricted to the O$(\Omega)$ seniority-zero subspace, as indicated in (\ref{redu}). This model has been a subject of sustained interest, both experimentally and theoretically, due to its ability to treat the competition between $T=0$ and $T=1$ pairing channels on an equal footing. The algebraic construction of the isovector plus isoscalar pairing Hamiltonian realized in the $LST$-coupling scheme was first proposed in \cite{Flowers1964FP18}, and its explicit matrix representation was derived in \cite{pang,Hecht1985FP20}. Subsequent analyses and applications of the O(8) model were carried out in \cite{Evans1981FP21,Dussel1986FP22,Engel1997FP23}, exploring its implications for pairing correlations, nuclear binding energies, and low-lying excitation spectra. An exact solution of the charge-independent mean-field Hamiltonian, including $l$-orbit-dependent single-particle energies together with isovector and isoscalar pairing interactions, was presented in \cite{LermaH2007FP24,Skrypnyk2012FP25}. In these solutions, the isovector and isoscalar pairing strengths were taken to be equal, thereby preserving the U(4) symmetry and allowing for an analytic treatment of the model.

Pan and collaborators subsequently extended the O(8) model in the $LST$-coupling scheme to the O$_{{\cal S}T}$(8) model \cite{Pan2020FP28}, where ${\cal S}$ denotes the pseudo-spin arising from a specific pseudo-orbital and pseudo-spin decomposition. In this extended framework, it is shown that the isovector ${\cal S} = 0$ and $T = 1$ pairs correspond exactly to the familiar $J = 0$ and $T = 1$ pairs, while the isoscalar ${\cal S} = 1$ pairs are realized as linear combinations of $J$-odd pairs. This construction enables the pairing Hamiltonian to be used for estimating both isovector and isoscalar pairing interactions within a unified algebraic framework. By fitting the model to selected low-lying $J = 0^+$ level energies of even-even and odd-odd nuclei with $A = 18$--$28$ up to the half-filling in the $sd$-shell, and comparing the results with those obtained from the original O(8) model, it was demonstrated that the isoscalar pairing interaction can be neglected in the lower-energy part of the spectra, insofar as binding energies and a few $J = 0^+$ excited levels of these nuclei are concerned.

With the mean-field plus isovector pairing interaction only, the contributions from nn, pp, and np pairing to the ground state or the lowest $J = 0^+$ state of these nuclei were estimated. It was found that the isovector np pairing contribution to the binding energy in odd-odd $N = Z$ nuclei is systematically larger than that in even-even nuclei, reflecting the enhanced role of isovector correlations when the valence proton and neutron numbers are both nonzero. Furthermore, the isoscalar np-pair content at the lowest $J = 0^+$ state of these nuclei was also evaluated. The results clearly show that the isoscalar pair content in the lowest $J = 0^+$ state of $N = Z$ and $N = Z \pm 2$ nuclei increases with the number of valence nucleons, particularly in even-even nuclei. These results distinguish the strength of the isoscalar pairing interaction from the isoscalar pair content of the eigenstates. A small contribution to the energy therefore need not imply weak isoscalar pairing correlations.

The same motivation led to an extension of the interacting boson model that accommodates both pairing channels. In \cite{Elliot1981FP26}, isovector $s$- and $d$-pairs with spin $S=0$ and isoscalar $s$- and $d$-pairs with spin $S=1$ were introduced as building blocks, giving rise to the so-called IBM-4. This model generalizes the earlier IBM-3, which included only isovector $s$- and $d$-pairs and thus neglected the isoscalar pairing channel \cite{Elliot1980FP27}. The explicit inclusion of both pairing types makes IBM-4 particularly well suited for describing nuclei in the $N \approx Z$ region, where $T=0$ and $T=1$ correlations are expected to coexist and compete.

The largest dynamical symmetry group of IBM-4 is $\mathrm{U}_{sdST}(36)$, which contains a rich spectrum of subgroup chains corresponding to various physical limits. These symmetry limits have been exploited extensively to interpret the structure of $N \approx Z$ nuclei. For example, odd--even staggering in $T=0$ pairs within the ground states of $N=Z$ nuclei was analyzed using the $\mathrm{U}_{sd}(6) \otimes \mathrm{U}_{ST}(6)$ and $\mathrm{O}_{sdST}(36)$ chains \cite{VanIsacker1997FP110}. In a related study, the ground $T=1$ band and the excited $T=0$ band of the self-conjugate nucleus $^{74}\mathrm{Rb}$ were interpreted within the $\mathrm{O}_{sdST}(36)$ limit, revealing the coexistence of different pairing regimes \cite{Kota1998FP114}. Furthermore, four-particle correlations were investigated in the $\mathrm{U}_{sST}(6) \oplus \mathrm{U}_{dST}(30)$ limit with $n_d = 0$, and their connection to the O(8) model was elucidated. 
Within IBM-4, Wigner's spin-isospin symmetry is equivalent to the $\mathrm{O}_{ST}(6)$ limit, underscoring its relevance for describing isoscalar pairing correlations.

The applicability of the IBM-4 symmetry limits is further motivated by empirical successes in neighboring regions. The O(6) limit of IBM-1 has long been recognized for its ability to describe $\gamma$-soft nuclei \cite{Iachello2011FP120}, while the $\mathrm{O}_{sd}(18) \supset \mathrm{O}_{d}(15) \oplus \mathrm{O}_{sT}(3)$ limit of IBM-3 has been shown to produce shape coexistence phenomena observed in some heavy $N=Z$ nuclei \cite{Kota1998FP117,Ginocchio1996FP121}. These observations suggest that the group chains descending from the $\mathrm{O}_{sdST}(36)$ limit of IBM-4 are particularly relevant for $N \approx Z$ nuclei near the proton drip line, with mass numbers $A \sim 60$--$100$ \cite{Kota2000FP122}. In this mass region, the competition between deformation, pairing, and shell effects gives rise to a variety of structural phenomena, making it a fertile ground for testing the predictions of the IBM-4 framework.

A simplified version of the model, restricted to the $L=0$ sector, was studied in Ref.~\cite{VanIsacker1998FP123}. While this reduced model cannot fully capture the entire physics content of the O(8) model \cite{Flowers1964FP18}, when moving away from the symmetry limits the discrepancies are found to be very small. This suggests that the essential physics of np pairing in the $L=0$ channel can be effectively described even in a truncated space. Consequently, despite these minor differences, both the full and simplified versions of the model can, in their own right, provide useful insights into the role of np pairing in $N \approx Z$ nuclei, each offering complementary perspectives on the underlying nuclear dynamics.


Beyond the IBM-4 framework, other theoretical approaches have been developed. In the end of the 1990s, Van Isacker, Juillet, and Nowacki introduced the pseudo-SU(4) model for heavy nuclei in the $pf$-shell \cite{Isacker1999PRL}, motivated by the concept of pseudo-spin symmetry \cite{Arima1969FP125,Hecht1969FP126,Bahri1992PRL,Draayer1993FP127}. In this framework, the physical spin is replaced by the pseudo-spin within a pseudo-orbital and pseudo-spin decomposition scheme. This substitution allows for a more efficient truncation of the nuclear shell model and provides an approximate symmetry that can describe GT transitions and other spin-dependent observables in medium-mass and heavy nuclei. The pseudo-SU(4) model has recently been applied to describe $\beta$-decay into self-conjugate $N=Z$ nuclei with mass numbers $58 \leq A \leq 70$ \cite{Isacker2023Symmetry}. Remarkably, it has been shown that a restoration of pseudo-SU(4) symmetry for $A = 70$ is supported by experimentally observed GT strength distributions, suggesting that this symmetry may be approximately realized in certain regions of the nuclear chart. The microscopic origin of pseudo-spin symmetry has continued to be investigated, including the chiral symmetry of massless hadrons \cite{Blokhin1995PRL}, the role of the tensor force~\cite{Nakada2024PRC}, and the emergence of pseudo-spin symmetry as the momentum resolution of the nuclear interaction is lowered~\cite{Ding2026PRL}.

More recently, following the proxy-SU(3) scheme  \cite{Bonatsos2017FP128,Bonatsos2023FP129}
in the $A = 60$--$90$ mass region, the proxy-SU(4) model has been developed \cite{Kota2024PScr}. This approach extends the proxy-SU(3) idea, which has been successful in describing deformation and collective motion, to include spin--isospin degrees of freedom. As pointed out in \cite{Kota2024PScr}, the proxy-SU(4) model opens new avenues for understanding the interplay between deformation and pairing in medium-mass nuclei. However, the authors also emphasize that further theoretical developments and experimental investigations are needed to fully assess the predictive power and limitations of this new symmetry scheme. In particular, systematic studies of GT transitions, electromagnetic properties, and nuclear matrix elements for double-beta decay would be valuable in testing the robustness of the proxy-SU(4) approach across different isotopic chains.

\subsection{The Nuclear Lattice Effective Field Theory}
\label{subsubsec:NLEFT}

In the era of \textit{ab initio} theory, Wigner's symmetry has also taken on a
computational role within nuclear lattice effective field theory (NLEFT), wherein nucleons are
propagated on a Euclidean spacetime lattice and their properties are extracted through Monte
Carlo simulation~\cite{Lahde2014PLB,Elhatisari2016PRL,Lee2025ARNPS}. In the lattice program
that exploits the symmetry, the interaction is built from the total nucleon density, summed
over all four spin-isospin components, so that it is left invariant by SU(4) rotations among
those components and therefore respects Wigner's symmetry~\cite{Lu2019PLB}. Such an
SU(4)-symmetric interaction is advantageous since it keeps the Monte Carlo weights
positive and thereby removes the fermion sign problem, exactly so for even-even
nuclei~\cite{Lu2019PLB,Niu2025PRL}; a related positivity of the Euclidean path integral in
the Wigner limit was established much earlier by Chen, Lee and Sch\"afer in deriving energy
inequalities for light nuclei~\cite{Chen2004PRL}. Interactions of this kind have been used
to describe the binding of nuclei up to the calcium region~\cite{Lu2019PLB}, the
thermodynamics of nuclear matter~\cite{Lu2020PRL}, and clustering in hot, dilute
matter~\cite{Ren2024PLB}, and they reproduce the low-lying spectrum of $^{12}$C, including
the Hoyle state~\cite{Shen2021EPJA}. More recently, the same lattice framework, now
employing high-fidelity chiral interactions, has been applied to the beryllium
isotopes~\cite{Shen2025PRL} and to the radii of the oxygen isotopes~\cite{Ren2025PRL}.

\section{\textit{Ab Initio} Nuclear Theory and Symmetry-Adapted Approaches}
\label{sec:abinitio}

\subsection{The \textit{Ab Initio} Landscape}
\label{subsec:shell-model}

At the onset of the 21st century, advances in $\chi$EFT 
and high-performance computing triggered a transformation in nuclear theory. 
A variety of \textit{ab initio} methods were introduced that share a common philosophy, 
whereby the nucleus is modeled in a large basis capturing all possible spatial, spin and isospin configurations 
of its constituent nucleons, which interact through an effective, yet realistic potential rooted in 
QCD~\cite{Stroberg2019ARNPS,Hergert2020FP}. Broadly, these methods fall into two families~\cite{Brown2025JPG}. 
The first comprises the quasi-exact approaches, which solve the many-body Schr\"odinger equation with a
controlled truncation error for light and, increasingly, medium-mass systems; these include
the NCSM, which diagonalizes the Hamiltonian in a flexible
harmonic-oscillator basis to deliver full spectra and associated wave
functions~\cite{Navratil2000PRL,Navratil2000PRC,Barrett2013PPNP},
its symmetry-adapted variants (namely, the SA-NCSM \cite{Dytrych2013PRL,Launey2016PPNP,Dytrych2020PRL},
the symplectic no-core configuration interaction (SpNCCI) framework \cite{McCoy2020PRL,McCoy2026},
and the SAM \cite{Dang2026}),
QMC \cite{Carlson2015RMP,Lynn2019ARNPS}, and
the NLEFT \cite{Lahde2014PLB,Elhatisari2016PRL,Lee2025ARNPS} discussed
above in Section~\ref{subsubsec:NLEFT}. The second comprises the more
approximate but more widely applicable polynomial-scaling methods, such as coupled-cluster (CC)
theory~\cite{Hagen2014RPP} and the in-medium similarity renormalization group 
(IMSRG) \cite{Hergert2016PR,Stroberg2019ARNPS} 
(and valence-space IMSRG \cite{Stroberg2021PRL}),
which trade explicit wave functions for soft scaling and thereby extend first-principles computations 
to heavy nuclei. For a broad survey of each method, see Refs.~\cite{Hergert2020FP,Brown2025JPG}.

\begin{figure}[!t]
\centering
\includegraphics[width=0.8\textwidth]{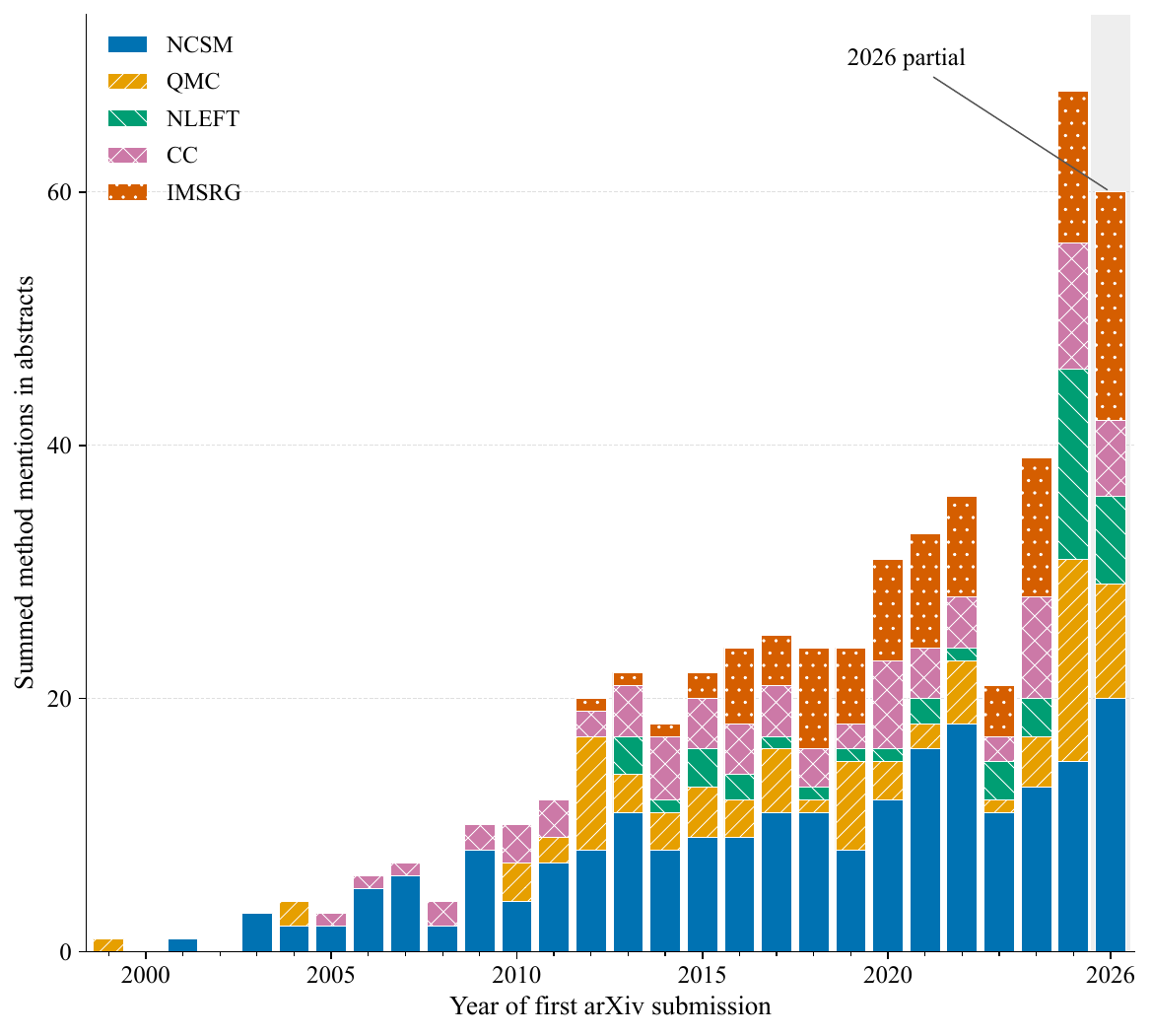}
\caption{Annual arXiv method mentions within the \textit{ab initio} nuclear-theory corpus,
grouped by the year of first submission, from a snapshot retrieved on
28 October 2026 using the \href{https://export.arxiv.org/api/query}{public arXiv API}.
The corpus subsumes 1,476 \texttt{nucl-th} records, including cross-listings, whose retrieved
abstracts contain ``\textit{ab initio}'' or ``first principles''; abstracts are
screened for the five method families by name and selected acronyms.
The stacked segments show the no-core shell model (NCSM, including SA-NCSM
and other extensions), quantum Monte Carlo (QMC), nuclear lattice
effective field theory (NLEFT), coupled cluster (CC), and the in-medium similarity
renormalization group (IMSRG). Each
arXiv identifier is counted once per method, irrespective of revisions. Papers mentioning
multiple methods contribute to multiple segments, so each bar gives a sum of
method mentions rather than a count of distinct papers. Mentions include reviews
and comparisons, and papers omitting the searched names may be missed. The shaded
2026 interval is a partial year through the retrieval date.}
\label{fig:arxiv-annual-counts}
\end{figure}

Figure \ref{fig:arxiv-annual-counts} displays a snapshot of annual method mentions within
the \textit{ab initio} nuclear-theory literature on arXiv as of 28 October 2026. The corpus
comprises 1,476 distinct records assigned to \texttt{nucl-th}, including cross-listings, whose
abstracts contain ``\textit{ab initio}'' or ``first principles,''
and within this corpus each abstract is screened for the five method families.
The annual counts provide a bibliometric view of the field;
method mentions can overlap and do not establish that a calculation uses the
method being mentioned.

More than half of these frameworks inherit the language of the nuclear shell model, in which the
strong force between nucleons is first replaced by an average field, reducing the $A$-body
problem to a set of nucleons moving independently in a three-dimensional harmonic oscillator (3DHO)
potential augmented by a spin-orbit term~\cite{Mayer1948PR,Mayer1949PR,Haxel1949PR}. The
resulting single-particle states
\begin{linenomath}
\begin{equation}
  \psi_{\eta lj m_j m_t} = R_{\eta l}(r)\left[Y_l(\theta,\phi)\times\sigma\right]^j_{m_j}\tau_{m_t},
\label{eq:shell-basis}
\end{equation}
\end{linenomath}
are labeled by the oscillator shell $\eta$, the orbital angular momentum $l$,
the total angular momentum $j$ and its projection $m_j$,
and the isospin projection $m_t$, with the spin-orbit term lifting the degeneracy
within each shell to reproduce the observed magic numbers. This independent-particle picture
cannot by itself capture the multi-nucleon correlations responsible for clustering and
deformation. Therefore, the interacting version was introduced to restore these correlations by freezing an inert
core and letting the valence nucleons interact through an effective residual interaction,
and a long line of work has sought to derive such interactions from the underlying
nucleon-nucleon force, with pioneering contributions by
Bertsch~\cite{Bertsch1965NP}, Kuo and Brown~\cite{Kuo1966NP}, and Kuo~\cite{Kuo1967NPA}.
See Ref. \cite{Stroberg2019ARNPS} for a more detailed review. 
In the \textit{ab initio} no-core settings, the core is removed and all nucleons are treated as active.

Although an $A$-nucleon system admits forces up to $A$-body rank, these are
hierarchically ordered, thus practical calculations retain two- and, at most, three-body terms. 
For simplicity, here we review a two-body Hamiltonian,
\begin{equation}
  \mathbf{H} = \frac{A-1}{A} \sum_{i}^A \frac{\mathbf{p}_i^2}{2m}
  + \sum_{i<j}^A \left[ V^{NN}_{ij} - \frac{1}{A} \frac{\mathbf{p}_i \mathbf{p}_j}{m}\right]
  = \sum_{i}^A \mathcal{K}_i + \sum_{i<j}^A \mathcal{V}^{NN}_{ij},
\end{equation}
where two effective operators have been introduced, $\mathcal{K}$ containing only the
one-body kinetic energy and $\mathcal{V}^{NN}$ subsuming both the two-body kinetic energy 
due to center-of-mass subtraction and the two-body nuclear potential $V^{NN}$. 
This formula, though simple, faces serious challenges stemming from the
indistinguishability of the nucleons. Therefore, a more convenient approach is to perform
calculations in the second quantization framework~\cite{Dytrych2008}, where the Hamiltonian
is evaluated as 
\begin{equation}
  \mathbf{H} = \sum_{\alpha\beta}\bra{\alpha}\mathcal{K}\ket{\beta}
  \mathbf{a}^\dagger_\alpha \mathbf{a}_\beta
  + \frac{1}{4}\sum_{\alpha\beta\gamma\delta}
  \bra{\alpha\beta}\mathcal{V}\ket{\gamma\delta}
  \mathbf{a}^\dagger_\alpha\mathbf{a}^\dagger_\beta\mathbf{a}_\delta\mathbf{a}_\gamma,
\label{eq:Hamil}
\end{equation}
Here, we introduce a short-hand notation $\ket{\alpha} := \ket{\eta l j m_j m_t}$;
$\mathbf{a}^\dagger_\alpha$ is the creation operator through which a particle pops out
from the vacuum to occupy a single-particle state ($\ket{\alpha} = \mathbf{a}^\dagger_\alpha \ket{0}$),
and $\mathbf{a}_\alpha$ is the annihilation operator which destroys the same particle
($\ket{0} = \mathbf{a}_\alpha \ket{\alpha}$). To encode antisymmetrization, these
operators obey the following anti-commutation relations \cite{Bahri1994CPC}
\begin{equation}
  \{ \mathbf{a}_\alpha, \mathbf{a}_\beta^\dagger \} = \delta_{\alpha\beta}
  \quad \text{and} \quad
  \{ \mathbf{a}_\alpha, \mathbf{a}_\beta \} = \{ \mathbf{a}_\alpha^\dagger, \mathbf{a}_\beta^\dagger \} = 0.
\end{equation}
The matrix elements---$\bra{\alpha} \mathcal{K} \ket{\beta}$ and antisymmetrized 
$ \bra{\alpha \beta} \mathcal{V} \ket{\gamma \delta}$---can be computed via a publicly accessible 
Fortran library called \texttt{NuHamil}~\cite{Miyagi2023EPJA}, 
which provides several realistic nuclear interactions.

What distinguishes the symmetry-adapted approaches (SA-NCSM, SpNCCI, SAM) within 
this \textit{ab initio} landscape is their organizing principle. Rather than treating the symmetries 
as an output to be measured after a calculation, they build the relevant emergent symmetries 
of the strong nuclear force directly into the many-body basis. This renders the symmetries not 
merely visible but exploitable, allowing the physically dominant configurations to be identified 
and retained, while the otherwise explosive dimensionality of the full NCSM space is brought under 
control. It is precisely this capability that positions the symmetry-adapted approaches to make Wigner's
supermultiplet symmetry computationally actionable, as developed in the remainder of this review.

\subsection{Elliott's SU(3) Symmetry and the SA-NCSM}

\subsubsection{SU(3) Classification of Nuclear States}
\label{subsec:su3}

Let us examine the algebraic structure of each major oscillator shell of the 3DHO Hamiltonian. 
For the $\eta$-th shell having degeneracy of
$4\Omega = 2(\eta + 1)(\eta + 2)$ (where $\Omega = (\eta + 1)(\eta + 2)/2$ is the number of
spatial orbitals in the shell, see Section \ref{subsec:apps}),
one can define creation and annihilation operators that move
particles within the same shell as $\mathbf{b}^\dagger_{\alpha i}$ and
$\mathbf{b}_{\alpha i}$ with $i=1,2,\ldots,\Omega$ and
$\alpha = m_s m_t = \uparrow\uparrow, \uparrow\downarrow, \downarrow\uparrow, \downarrow\downarrow$.
The biproducts of these operators form a closed set of $\Omega^2$ generators of the unitary
group U($\Omega$),
\begin{linenomath}
\begin{equation}
  E_{ij} = \sum_\alpha \mathbf{b}^\dagger_{\alpha i} \mathbf{b}_{\alpha j},
\end{equation}
\end{linenomath}
which satisfy the following conjugation and commutation relations
\begin{linenomath}
\begin{equation}
  E_{ji} = E_{ij}^\dagger \quad \text{and} \quad [E_{ij},E_{kl}] = \delta_{jk} E_{il} - \delta_{il} E_{kj}.
\end{equation}
\end{linenomath}
The action of the generator $E_{ij}$ can be understood as moving a particle to the $i$-th
direction from the $j$-th direction in the $\Omega$-dimensional space of the oscillator shell.
A multi-particle configuration within the $\eta$-th shell can be represented by a U($\Omega$)
irrep, $[n_{1\Omega} \ge n_{2\Omega} \ge \ldots \ge n_{\Omega\Omega}]$. Since the spin-isospin label $\alpha$
can assume one of the four combinations, each partition $n_{i\Omega}$ of the U($\Omega$) irrep ranges
between 0 and 4.

If we redefine the creation and annihilation operators with $i=x,y,z$ and $\alpha=m_s m_t$,
their biproducts form a closed set of 9 generators of the U(3) group, whose irreps contain
three quantum numbers $[n_{13} \ge n_{23} \ge n_{33}]$. In this case, one interprets the
action of the generator $E_{ij}$ as shifting a particle from the $j$-th direction to the
$i$-th direction in the real three-dimensional space, while preserving its spin and isospin
quantum numbers. The U(3) group is a subgroup of the U($\Omega$) group, thus there exists a
decomposition from a single U($\Omega$) irrep to multiple U(3) ones, each of which may appear
multiple times in the decomposition. Draayer \textit{et al.}~\cite{Draayer1989CPC} were the
first to advance a computer program written in Fortran that executes this decomposition
procedure. Recently, Langr \textit{et al.}\ have produced modern C++ libraries for
large-scale shell model calculations in the proton-neutron formalism~\cite{Langr2019CPC} and
in the isospin formalism~\cite{Langr2025}; the latter is released as the open-source library \texttt{untou3},
which is part of the \texttt{DeepSANE} ecosystem.

It is possible to invoke a linear mapping of the nine generators of the U(3) group to a new
set of operators~\cite{Pan2016NPA}
\begin{linenomath}
\begin{align}
  \mathcal{C}_1 &= \sum_{i=1}^3 E_{ii} \\
  L_0 &= E_{11} - E_{22} \text{, } L_+ = (L_-)^\dagger = \sqrt{2} (E_{13} + E_{32}) \\
  Q_0 &= \frac{E_{11} + E_{22} - 2E_{33}}{\sqrt{6}} \text{, }
  Q_1 = (Q_{-1})^\dagger = \frac{E_{32} - E_{13}}{\sqrt{2}} \text{, }
  Q_2 = (Q_{-2})^\dagger = E_{12}.
\end{align}
\end{linenomath}
Here $\mathcal{C}_1$ is the number operator which counts the total quanta of the irrep
(i.e., $\langle \mathcal{C}_1 \rangle = n_{13} + n_{23} + n_{33}$), $\{L_+, L_0, L_-\}$ form
the subgroup SO(3), and $Q_\mu$'s ($\mu=0, \pm 1, \pm 2$) are the tensor components of the
``algebraic'' mass quadrupole operator, which do not form a group on their own. 
By removing the number operator $\mathcal{C}_1$, one
is left with a set of eight generators which form a special unitary subgroup SU(3), whose
irreps have two non-negative quantum numbers $(\lambda,\mu)$. Ignoring the spin and isospin
degrees of freedom, the states of a nucleus restricted to a single shell can be classified
by quantum numbers of the group-subgroup chain SU(3) $\supset$ SO(3) $\supset$ SO(2) as
$\ket{(\lambda,\mu)\kappa L M}$, where $L$ is the orbital angular momentum with projection
$M$, and $\kappa$ is a multiplicity label to keep track of the recurrence of the same $L$
within a single SU(3) irrep. 

\subsubsection{A Brief History of SU(3) in Nuclear Structure}

The adoption of the SU(3) symmetry in modeling nuclear structure has a long history,
stretched back to the late 1950s with the foundational papers of Elliott \cite{Elliott1958PRSA,Elliott1958PRSA-II}, 
where the classification scheme of nuclear states according to representation of the SU(3) group 
shown previously was first introduced. Moving into the 60s, Elliott's SU(3) model gained limited
momentum due to a lack of scalable computational tools that could produce coupling and
recoupling coefficients for this three-dimensional group. It was not until 1973 that Draayer and
Akiyama advanced a pair of papers, the first of which laid out the theoretical foundation of
a generic algorithm for computing these coefficients, and the second complementary one
proffered the first Fortran implementation of this algorithm~\cite{Draayer1973JMP,Akiyama1973CPC}.

The introduction of the so-called \texttt{SU3} package by Draayer and Akiyama paved the way
for applications of the SU(3) symmetry in nuclear modeling to take off during the following
decades. Started in the 70s and 80s were the symplectic shell model developed by Rosensteel
and Rowe~\cite{Rosensteel1977AnnP,Rosensteel1980AnnP,Rowe1985RPP} and the pseudo-SU(3) model
for heavy and deformed nuclei~\cite{Raju1973NPA}. During this period, Draayer and associates
also developed symmetry-grounded nuclear potentials going beyond the traditional
quadrupole-quadrupole interaction that was known to reproduce rotational spectra very well.
These algebraic interactions are built from the SU(3) generators and Casimir invariants of
the group; and they are referred to as ``the Integrity
Basis''~\cite{Gilmore1985JMP,Draayer1985NPA}. The culmination of this development was the
interpretation of collective behavior from shell-model perspectives. Specifically, by mapping
the quadratic and cubic powers of the mass quadrupole moment to the SU(3) Casimir operators,
Draayer \textit{et al.}\ discovered a very simple relationship between the quantum numbers
$(\lambda,\mu)$ and the deformation parameters $(\beta,\gamma)$ of the liquid drop
model~\cite{Castanos1988ZPA,Draayer1989PRL} as follows
\begin{linenomath}
\begin{equation}
  \beta^2 = \frac{4\pi}{5(A\langle r^2\rangle)^2}\!\left(\lambda^2+\lambda\mu+\mu^2+3\lambda+3\mu+3\right),
  \quad
  \gamma = \arctan\!\left[\frac{\sqrt{3}(\mu+1)}{2\lambda+\mu+3}\right],
\end{equation}
\end{linenomath}
where $A$ is the number of nucleons and $\langle r^2\rangle$ is the mean squared mass radius.
This mapping, for the first time, made it possible to infer the intrinsic shape of a nucleus
from a shell-model description, 
particularly
\begin{itemize}
\item $\lambda = \mu = 0$, then the irrep describes a spherical shape,
\item $\lambda > 0, \mu = 0$, then $0 \le \gamma \le \pi/6$ corresponds to a prolate shape,
\item $\lambda = 0, \mu > 0$, then $\pi/6 \le \gamma \le \pi/3$ corresponds to an oblate shape.
\end{itemize}
Note that the shape variables $(\beta,\gamma)$ can vary continuously without restriction,
whereas the SU(3) quantum numbers assume discrete values that are restricted to a finite set.
This can be explained by the fact that the liquid drop model does not take into account the
fermionic nature of nuclei~\cite{Castanos1988ZPA}. Inversely, that the set of SU(3) irreps is
finite implies the existence of a finite set of $(\beta,\gamma)$ values that can satisfy the
particle nature of the nucleons \cite{Castanos1988ZPA}.

During the last ten years of the 20th century, progress continued to be made with an
important shift from a seemingly bosonic picture where nuclear states are classified with
only spatial degrees of freedom to a more realistic picture with spin symmetry started to be
built into the Hilbert space and operator structure through the tensorial second quantization
formalism~\cite{Bahri1994CPC,Escher1998JMP}. This transformation was largely enabled by the
capability of high-performance computer infrastructure and marked the first step toward what
would come during the next thirty years.

\subsubsection{The Symmetry-Adapted No-Core Shell Model}

Transitioning into the 21st century, while algebraic models continued to be developed and applied
(for instance, the pseudo-$\rm SU(3)$ \cite{Kota2020,Cseh2020PRC,Gade2025NatPhys}, 
proxy-$\rm SU(3)$ \cite{Bonatsos2017FP128,Bonatsos2023FP129,Cseh2020PRC},
and quasi-$\rm SU(3)$ \cite{Zuker1995PRC,Gade2025NatPhys} schemes, as well as the
multiconfigurational dynamical symmetry (MUSY) \cite{Cseh2021PRC}),
an ambitious program was initiated at Louisiana State University and led by Draayer. This project came at the center
of three important technological and scientific milestones: i) supercomputers with parallel
processing capability became ever more powerful, ii) the NCSM was introduced by
Navr\'atil, Vary and Barrett~\cite{Navratil2000PRL,Navratil2000PRC}, and iii) the work of
Bahri and Draayer \cite{Bahri1994CPC} made it possible to compute matrix elements of any two-body
operator in an SU(3)$\otimes$SU(2)-adapted basis through the second quantization formalism.
These advancements led to the creation of what is now known as the \textit{ab initio}
Symmetry-Adapted No-Core Shell Model (SA-NCSM)\cite{Dytrych2007PRL,Dytrych2007PRC,Dytrych2008JPG1,Dytrych2008JPG2,Dytrych2013PRL,Dytrych2020PRL,Launey2016PPNP}
with Dytrych and Launey as the driving force. The most impactful
discovery enabled by the SA-NCSM was demonstration of a natural emergence of an orderly
symmetry pattern in structure of atomic nuclei from realistic nuclear potentials;
specifically, nuclear wave functions are dominated by only a few SU(3) irreps, hence only a
few static shapes. Mastering the knowledge of such a feature,
the inventors of the SA-NCSM devised a strategy to squeeze the explosive NCSM space to only a
set of tractable configurations that are vital for nuclear dynamics. To date, the SA-NCSM
continues to serve as a premier model for probing fundamental physics and nuclear
reactions~\cite{Launey2021AnnRev,Launey2023SciPost,Launey2026PPNP}.

The SA-NCSM established the spatial analogue of the story in this review: a symmetry
emergent from realistic interactions that, once identified and encoded in the basis,
makes large-scale calculations tractable. The SAM completes this picture by adding the
spin-isospin U(4) counterpart.

\section{Coupling and Recoupling Coefficients: Enabling Technology}
\label{sec:coefficients}

A rigorous exploitation of Wigner's symmetry in nuclear structure modeling requires 
the ability to compute (Wigner) coupling and recoupling (Racah) coefficients, 
which enter every operator matrix element via the Wigner--Eckart theorem 
and constituted a principal computational bottleneck for decades.
We recently addressed this challenging problem for the $\rm U(4)$ group through a
modernized resolution for the $\rm U(3)$ group, which, we argue, is one of the enabling 
advances that has brought in a new symmetry-adapted approach to \textit{ab initio}
nuclear theory by converting Wigner's symmetry from a concept into a practical 
computational tool, and it is the technical foundation on which the SAM (Section~\ref{sec:SAM}) is built.
 
\subsection{U(3) Coupling and Recoupling Coefficients}

The first computational tool for the SU(3) $\supset$ SO(3) coupling coefficients that was 
first advanced in Fortran by Draayer and Akiyama in 1973~\cite{Draayer1973JMP,Akiyama1973CPC} 
relied on hypergeometric functions that involve large factorials and alternating sums. 
Along the way, various upgrades to the algorithm to improve
numerical precision with modern programming languages were introduced 
\cite{Bahri2004CPC,Dytrych2021CPC,Herko2026EPJA},
providing the engine for algebraic nuclear models.
However, this analytical formulation had remained unchanged for five decades.

In 2024, Dang \textit{et al.}~\cite{Dang2024EPJP1} introduced a new algorithm based on
numerical nullspace solvers and sidestepping the need for analytical formulas.
The key insight is to recast both the basis transformation (from the physical
SU(3) $\supset$ SO(3) basis to the canonical SU(3) $\supset$ SU(2)$\otimes$U(1) basis)
and the coupling coefficient evaluation as shared nullspace problems; specifically,
the physical basis states are found from the nullspace of $L_+$~\cite{Pan2016NPA},
and the canonical coupling coefficients from the shared nullspace of the canonical raising generators
$E_{12}$ and $E_{23}$, following the generic formulation of Alex \textit{et al.}~\cite{Alex2011JMP} for any unitary group.
This approach bypasses large factorials entirely, resolving the accuracy problem and
enabling the modern \texttt{u3lib} library~\cite{u3lib}, which is
equipped with parallel-processing capability and an AI-ready skill catalog for agentic orchestration.

\subsection{U(4) Coupling and Recoupling Coefficients}

An analogous challenge existed for U(4), but for a longer period of almost ninety years 
since Wigner's proposal of the theory. In spite of partial results that had been obtained 
by Hecht and Pang~\cite{Hecht1969JMP}, Draayer~\cite{Draayer1970JMP}, Partensky 
and Maguin~\cite{Partensky1978JMP}, and Rowe and Repka~\cite{Rowe1997FoP}, 
no generic algorithm and computational tool existed for arbitrary U(4) coupling and recoupling 
coefficients in the physical chain $\text{U}(4) \supset \text{SU}_S(2) \otimes \text{SU}_T(2)$.

Recent development of the new $\rm U(3)$ methodology and elementary coupling
coefficients by Pan \textit{et al.} \cite{Pan2023EPJP,Pan2023NPA,Pan2024CPC}
led Dang \textit{et al.}~\cite{Dang2024EPJP2} to build a sibling of \texttt{u3lib},
the library \texttt{u4lib}~\cite{u4lib}, which is written in C++ with Python
binder and provides parallel-processing and agent-ready capabilities. In addition to 
the coupling coefficients that are computed for the two group-subgroup chains, 
$\text{U}(4) \supset \text{SU}_S(2) \otimes \text{SU}_T(2)$
and $\text{U}(4) \supset \text{U}(3) \supset \text{U}(2) \supset \text{U}(1)$ from the nullspace
of the U(4) raising generators \cite{Dang2024EPJP2} and the recoupling coefficients that follow,
this library also supplies other algebraic tools, such as Casimir invariants, basis enumeration, 
inner and outer multiplicities, basis transformation, etc., making \texttt{u4lib} a complete
computational framework for the $\rm U(4)$ symmetry, suitable for a variety of applications.


Together, \texttt{u3lib} and \texttt{u4lib} resolve the central mathematical obstacles
to simultaneously encoding spatial U(3) and spin-isospin U(4) in an \textit{ab initio}
model, laying the foundation of the SAM presented in the next section. Both libraries are
released openly as part of the \texttt{DeepSANE} ecosystem.

\section{The \textit{Ab Initio} Symmetry Adapted Model}
\label{sec:SAM}

\subsection{Many-Body Basis}
\label{subsec:basis}

The \textit{ab initio} Symmetry Adapted Model (SAM) follows in the footsteps of its
predecessor---the SA-NCSM---by employing a large-scale microscopic model space that subsumes
all possible configurations of a nucleus and letting the constituent nucleons interact via a
realistic nuclear potential. Its developers chose the new name Symmetry Adapted Model (SAM) for two reasons. 
First, the SAM is a new \textit{ab initio} code built on the first general high-performance engines 
for U(4)~\cite{Dang2024EPJP2,u4lib} and U(3)~\cite{Dang2024EPJP1,u3lib} coupling and recoupling coefficients, 
which together enable a U(3)$\otimes$U(4)-adapted basis and selection on either symmetry or both. 
Second, the name conveys the broader aim of a flexible framework that covers high-quality nuclear forces,
state-of-the-art SRG methods, and many-body bases guided by emergent orderly patterns. The SAM is not
distinct from the SA-NCSM but inherits its basis-truncation concept while incorporating
further \textit{ab initio} techniques. Like the coefficient libraries on which it builds, 
the SAM's source code is part of the \texttt{DeepSANE} ecosystem.

\begin{figure}[t]
\centering
\includegraphics[width=0.8\textwidth]{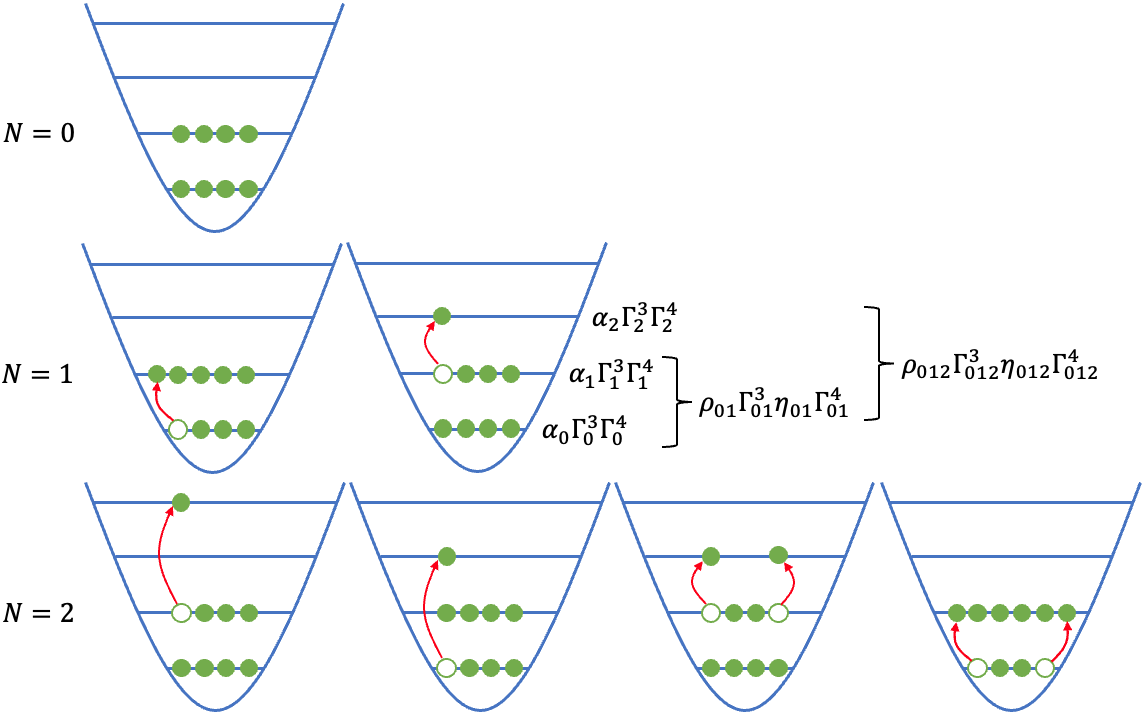}
\caption{Schematic demonstration of model space construction. The first row displays the
$0\hbar\omega$ space where all nucleons reside in the lowest shells. The second row
illustrates the $1\hbar\omega$ space where one nucleon is lifted to the next available shell.
The last row depicts the $2\hbar\omega$ space in which either one nucleon is raised by two
shells or two nucleons are elevated by one shell. The end of the middle row describes the
intershell coupling procedure.}
\label{fig:multi-shell}
\end{figure}

The Hilbert space is constructed similarly to the model space of
the NCSM and the SA-NCSM. Specifically, starting from the valence space
where all nucleons occupy the lowest shells, an amount of energy equal to the shell separation
energy $\hbar\omega$ is added to raise one nucleon up to the next available shell, generating the
$1\hbar\omega$ space. Similarly, one can generate the $2\hbar\omega$ space by raising either
one nucleon by two shells or two nucleons by one shell. The procedure can be repeated, see
Figure~\ref{fig:multi-shell}, and the total model space is defined by a parameter $N_{\max}$,
which signifies the maximum number of quanta that can be added to the system. For even
$N_{\max}$, the wave functions converge to states of natural parity, while states of opposite
parity reside within odd $N_{\max}$ model spaces. It is also worth noting that most nuclear
potentials do not mix parity, therefore, the subspaces that contribute to the total
$N_{\max}$ space differ by $2\hbar\omega$; in other words, the active subspaces are
$N_{\max},N_{\max}-2, N_{\max}-4, \ldots, 0$ or $1$.

Basis states are constructed by explicitly encoding both U(3) and U(4) symmetries.
For each $\eta$-th shell (with degeneracy $4\Omega = 2(\eta+1)(\eta+2)$), one identifies all
U($\Omega$) irreps representing the spatial degrees of freedom, then generates the conjugate
U(4) irreps denoted by $\Gamma^4 := [n_{14},n_{24},n_{34},n_{44}]$ to enforce
antisymmetrization. A decomposition from U($\Omega$) to U(3) irreps denoted by 
$\Gamma^3 := [n_{13},n_{23},n_{33}]$ is performed~\cite{Draayer1989CPC,Langr2019CPC,Langr2025},
with an additional multiplicity $\alpha$ tracking multiple recurrences of the same U(3)
irrep within each U($\Omega$) irrep. This procedure is repeated for all shells.

Once completed for each shell individually, intershell coupling takes place recursively, starting
from the lowest shell, as depicted in Figure~\ref{fig:multi-shell}. During coupling,
two multiplicities are required, namely $\rho$ for U(3) irreps \cite{Dang2024EPJP1} and 
$\eta$ for U(4) irreps \cite{Dang2024EPJP2},
arising from coupling of two irreps. (Note that this
multiplicity label $\eta$ should not be confused with the shell number.) Once all
intershell U(3) and U(4) irreps are obtained, two parallel decomposition algorithms
are run: i) each U(3) irrep is decomposed into orbital angular momenta $L$ with
multiplicity $\kappa$ \cite{Dang2024EPJP1}; ii) each U(4) irrep is decomposed into 
spin $S$ and isospin $T$ pairs with multiplicity $\zeta$ \cite{Dang2024EPJP2}. 
Two physical constraints are imposed, including the angular
momentum coupling condition $|L-S| \le J \le L+S$, and 
$T \ge |M_T| = | \text{neutron number} - \text{proton number}| /2 $.
The complete basis label is
\begin{linenomath}
\begin{equation}
\begin{aligned}
  \Big|\, \underbrace{(\alpha_0\Gamma^3_0\Gamma^4_0)(\alpha_1\Gamma^3_1\Gamma^4_1)...(\alpha_n\Gamma^3_n\Gamma^4_n)}_{\text{intrashell}};
  \underbrace{\rho_{01}\Gamma^3_{01}\eta_{01}\Gamma^4_{01}\,...\,\rho_{0...n}\Gamma^3_{0...n}\eta_{0...n}\,\Gamma^4_{0...n}}_{\text{intershell}};\,
  \kappa L;\,\zeta S T;\,J \,\Big\rangle,
\end{aligned}
\label{eq:state}
\end{equation}
\end{linenomath}
where $n$ denotes the highest occupied shell, the intrashell factors
$(\alpha_i\Gamma^3_i\Gamma^4_i)$ are the U(3) and U(4) irreps of the $i$-th shell, and each
intershell factor $\rho_{0\cdots i}\,\Gamma^3_{0\cdots i}\,\eta_{0\cdots i}\,\Gamma^4_{0\cdots i}$
is the cumulative irrep obtained when the $i$-th shell is coupled onto the already-coupled
shells $0,\ldots,i-1$, as illustrated in Figure~\ref{fig:multi-shell}. The last such coupling
defines the total irreps $\Gamma^3_{0...n}$ and
$\Gamma^4_{0...n}$, which are reduced to $\kappa L$ and $\zeta S T$ and
combined into the total angular momentum $J$.

\subsection{Hamiltonian in U(3)$\otimes$U(4) Tensor Form}
\label{subsec:Hamil}

Since the SAM works in symmetry-adapted bases, the creation and annihilation operators are
recast as good U(3)$\otimes$U(4) tensors---carrying the tensorial labels
$[\eta,0,0][1,0,0,0]$ and $[\eta,\eta,0][1,1,1,0]$, respectively---and the Hamiltonian of
Eq.~(\ref{eq:Hamil}) is expanded over a set of one- and two-body U(3)$\otimes$U(4) tensors.
This expansion was derived by extending the SU(3)$\otimes$SU(2) proton-neutron tensor 
transformation of the SA-NCSM \cite{Oberhuber2021DCDS} and is
presented in Supplemental Material of Ref. \cite{Dang2026}.

This tensor form is also advantageous as it allows the Wigner--Eckart theorem \cite{Wigner1993} 
to be invoked. Once the Hamiltonian is written in U(3)$\otimes$U(4) tensors, every
matrix element factorizes into a geometric part, built entirely from the U(3) and U(4) coupling coefficients,
and a dynamical part, the reduced matrix elements (RMEs), which are independent of the orbital,
spin, isospin, and total-angular-momentum quantum numbers. The RMEs can therefore be evaluated
once, in the convenient canonical group chain
$[\text{U}(3)\supset\text{U}(2)\supset\text{U}(1)]\otimes[\text{U}(4)\supset\text{U}(3)\supset\text{U}(2)\supset\text{U}(1)]$,
which is free of inner multiplicities ($\kappa$ and $\zeta$), and then reused to assemble all remaining matrix elements
from the coupling coefficients alone.

\begin{figure}[t]
\centering
\includegraphics[width=0.8\textwidth]{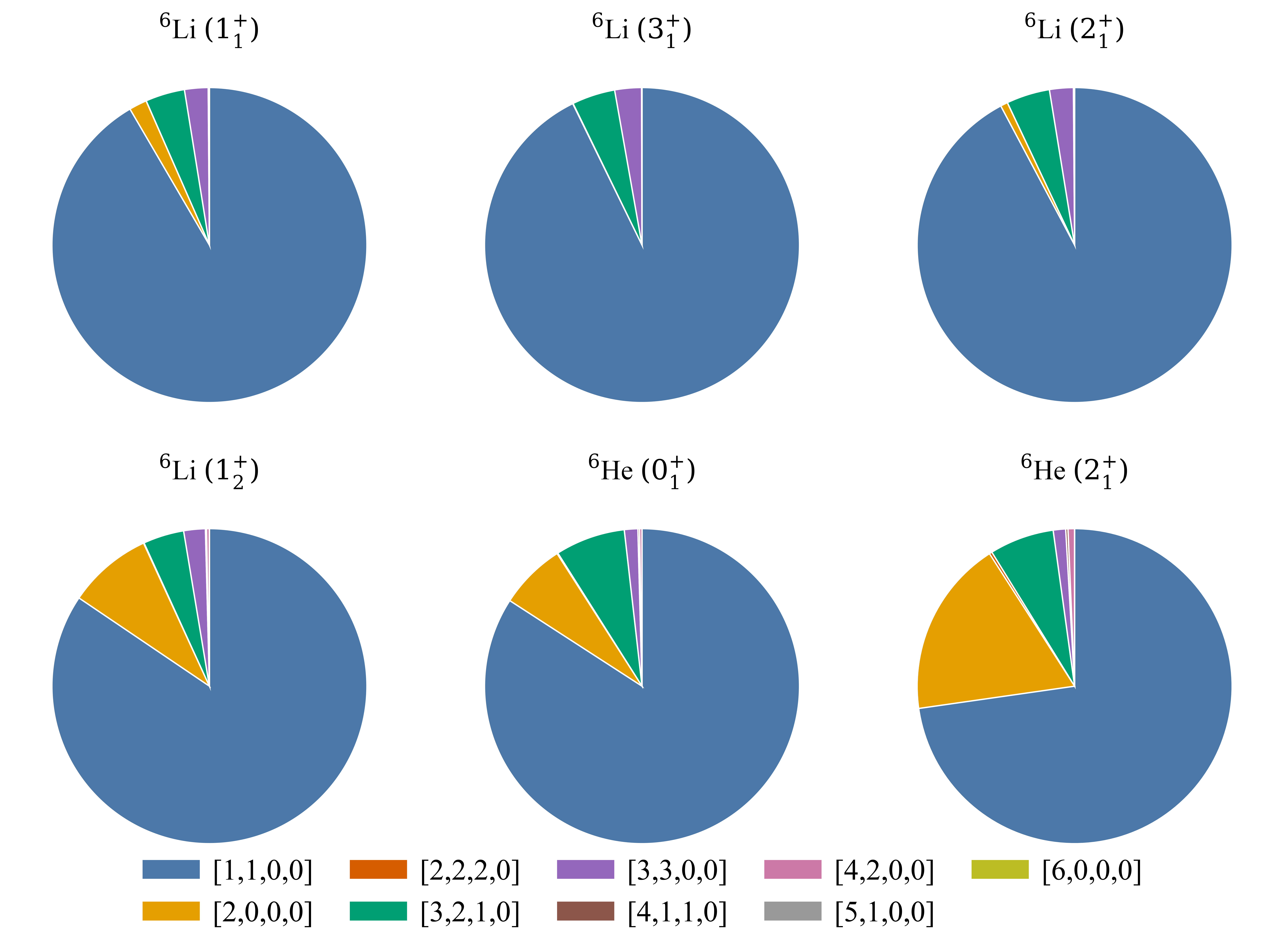}
\caption{Total U(4) irrep probabilities of the low-lying states of $^{6}$Li and $^{6}$He,
computed with the SAM using the N2LO$_{\rm opt}$ interaction in an $N_{\max}=10$ model
space. Each pie chart shows the probability of every U(4) irrep summed over all
$\hbar\omega$ subspaces. Data are taken from Ref. \cite{Dang2026}. The irreps are normalized
according to Ref. \cite{Dang2024EPJP2}.}
\label{fig:u4_pies}
\end{figure}

\subsection{First-Principles Probes Enabled by the SAM}
\label{subsec:results}

Beyond serving as a computational framework for nuclear structure, the SAM opens 
three avenues for examining Wigner's symmetry \cite{Dang2026}. First, the U(3)$\otimes$U(4) tensor expansion
can expose the symmetry content carried by a given interaction---a diagnostic that is not readily 
available in other many-body bases. Second, because the basis is built explicitly on U(4) irreps, 
the SAM resolves the supermultiplet structure of computed wave functions, making it possible 
to ask which spin-isospin configurations dominate nuclear binding. Third, once the dominant
irreps are determined, one can perform model space truncation for few- and many-body calculations
with less computational cost than full NCSM spaces, which is only possible with a symmetry-adapted basis.

A many-body basis constructed directly on U(4) irreps also carries an advantage over 
former approaches that infer the supermultiplet content by diagonalizing the quadratic
Casimir invariant $\mathcal{C}_2$ \cite{Vogel1993PRC,Muli2026PRL}. Resolving
the irreps explicitly exposes how much each contributes within every $\hbar\omega$ subspace
rather than an aggregate over the whole space that would be delivered by diagonalizing only $\mathcal{C}_2$,
making it possible to perform irrep selection at each $\hbar\omega$ subspace. 
Figure~\ref{fig:u4_pies} demonstrates the supermultiplet content for the low-lying states of $^{6}$Li and
$^{6}$He. Each state is overwhelmingly governed by a single leading U(4) irrep, $[1,1,0,0]$,
which carries between roughly $73\%$ and $93\%$ of the probability and is a direct
manifestation of Wigner's symmetry in the wave functions. This irrep also has the lowest eigenvalue of
$\mathcal{C}_2$, which is equal to $\mathbf{S}^2 + \mathbf{T}^2 + \mathbf{GT}^2$ in the physical chain
$\rm U(4) \supset SU_S(2) \otimes SU_T(2)$ (Section~\ref{subsec:group_theory}); hence, it might be interpreted
as suppression of spin-isospin polarizability of the nuclear force. This phenomenon was also found
to take place in infinite nuclear matter at low densities \cite{Isayev2006PRC}.
It was also found that U(4) irrep selection for these nuclei compresses the full NCSM space by at most half,
thus future structure calculations would benefit more from combined selection of both U(3) and U(4) irreps---as 
enabled by the SAM basis---to achieve more significant reduction of the model space.

Additionally, a U(4)-adapted basis separates conjugate irreps that share the same value of $\mathcal{C}_2$, 
which would otherwise be indistinguishable. In the case of $^{6}$Li and $^{6}$He, the conjugate
irreps $[2,0,0,0]$ and $[2,2,2,0]$, which are degenerate in $\mathcal{C}_2$, enter with markedly
different weights, the former exceeding the latter by factors ranging from about five to more
than one hundred across these states, see Figure~\ref{fig:u4_pies}. A probe based on the Casimir eigenvalues alone cannot
distinguish the two and may converge numerically to a mixture of them, whereas the SAM  
treats them separately by construction. It is also worth mentioning that the two conjugate irreps carry
different spin-isospin polarizability modes, and by extension, different physics. An analog
can be found in the SU(3) case; specifically, for the two conjugate irreps $(\lambda\mu)$
and $(\mu\lambda)$ with $\lambda > \mu$, which share identical quadratic Casimir eigenvalue, the former
corresponds to a prolate shape, whereas the latter corresponds to an oblate one.

\subsection{Machine Learning Nuclear Masses}

Building on the previous works that leverage Wigner's symmetry to model nuclear masses through the
Casimir operators \cite{Cauvin1981NPA,Burdet1968NuovoCim,Gaponov1982NPA,Isacker1997FoP}
and on modern machine-learning techniques \cite{Boehnlein2022RMP}, Dang \textit{et al.} recently 
adopted neural networks to fit the binding energies of known nuclei and predict extremely neutron-rich isotopes 
that are important for the rapid neutron capture (the $r$-process) responsible for nucleosynthesis of 
many elements heavier than iron in explosive astrophysical environment \cite{Dang2026_2}. In lieu of 
feeding the SU(3) and SU(4) Casimir operators directly as features like other approaches 
\cite{Li2024PLB,Utama2016PRC,Gao2021NST,Mumpower2022PRC,Huang2025PRC,Kim2026PRC,Lu2026PRC,Liu2025PRC,Lu2025PRC,Niu2018PLB,Utama2018PRC,Qu2025CPC,Kejzlar2023SciRep,Yuksel2024PRC,Ye2025PRC,Liu2026PRC,Jalili2026JPG}, 
the authors proposed a neural architecture that learns the coefficients $\{a,b,c,d,e\}$ as functions of 
neutron and proton numbers in the following bilinear formula
\begin{equation}
\label{eq:WINN}
\Delta BE = a N_{\hbar\omega} + b \mathcal{C}_2[{\rm SU(3)}] + c \mathcal{C}_3[{\rm SU(3)}]
		+ d \mathcal{C}_2[{\rm SU(4)}] + e \mathcal{C}_4[{\rm SU(4)}],
\end{equation}
where $\Delta BE$ is the residual after subtraction of the liquid-drop-model binding energy
and $N_{\hbar\omega}$ is the total number of harmonic-oscillator quanta. The Casimir
operators are substituted by their eigenvalues evaluated in the leading irreps.

This seemingly rigid bilinear form helps constrain the hypothesis space of the neural network without 
trading accuracy and extrapolation capability. Specifically, the authors benchmarked this model against 
two standard approaches---namely multilayer-perceptron network taking the operators as input features and its 
Gaussian variant---and found that this architecture design degrades most gracefully as data are withdrawn 
from the training set and yields more stable extrapolation (manifested by more consistent single-neutron 
separation energy prediction and elemental abundances of a binary neutron star merger). More importantly, 
this neural network is intrinsically interpretable without any post-hoc interpretability tools (e.g., the Shapley 
additive explanation) since each term's contribution to Eq. (\ref{eq:WINN}) can be read off directly after training.
Note that the leading irreps are chosen by hypothesizing that the lowest-$\mathcal{C}_2[{\rm SU(4)}]$ 
observation in realistic SAM wave functions for light nuclei \cite{Dang2026} extends to the entire nuclear chart,
including where the SU(4) symmetry is severely broken; the authors suggest that
the neural network might have addressed symmetry breaking and that irrep selection based on 
algebraic extensions of the SU(4) model (see Section~\ref{subsec:apps}) may improve model performance 
for future work \cite{Dang2026_2}.

\section{Conclusion and Future Outlook}
\label{sec:conclusions}

\begin{figure}[h!]
\centering
\includegraphics[width=0.5\textwidth]{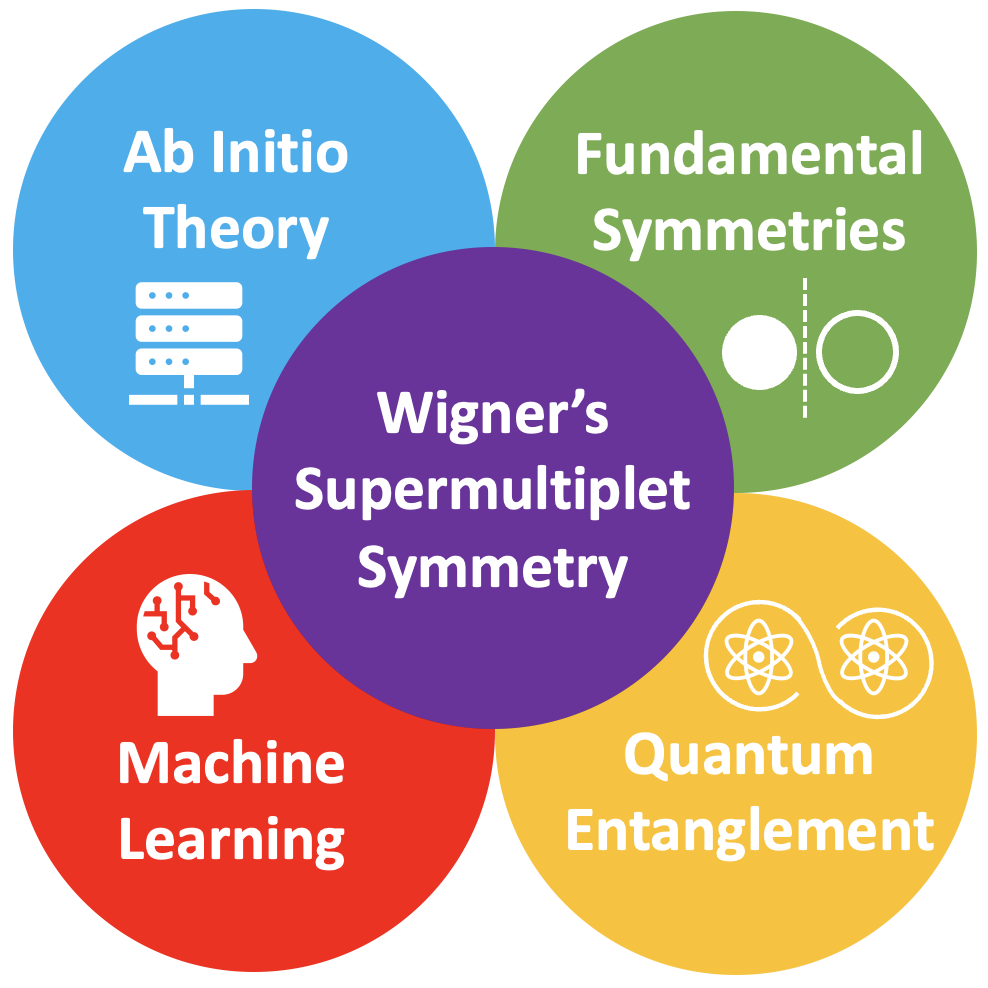}
\caption{Synopsis of the perspectives through which this review examines Wigner's
supermultiplet symmetry (center). The four surrounding themes recur throughout the text ---
its \textit{ab initio} realization in the SAM, its standing among the fundamental symmetries
of the strong interaction and the weak processes, the quantum-entanglement
arguments that account for its emergence, and its role in machine-learning descriptions of
nuclear observables.}
\label{fig:sypnosis}
\end{figure}

This review has traced the journey of Wigner's supermultiplet symmetry from its 1937 proposal to its
present-day \textit{ab initio} realization in the SAM, examining it through the complementary
lenses of \textit{ab initio} theory, fundamental symmetries, quantum entanglement, and
machine learning that are collected in Figure~\ref{fig:sypnosis}. The symmetry emerges robustly from
low-energy two-nucleon scattering data, from the large-$N_c$ limit of QCD, from nuclear
lattice calculations, and from entanglement-suppression arguments---all of which evolve along 
converging lines of evidence that the strong nuclear force is approximately spin-isospin invariant. After
decades of oscillating interest, driven first by the success of the spin-orbit shell model
and later revived by the observation of isobaric analog states, the symmetry is now probed
directly from realistic, large-scale nuclear wave functions for the first time. The enabling
step was the development of nullspace-based algorithms for U(3) and U(4) coupling and
recoupling coefficients, which resolved a computational obstacle that had stood for nearly
nine decades. Built on these coefficients, the SAM shows that realistic chiral interactions
are strongly U(4)-symmetric, that nuclear wave functions concentrate in a handful of
supermultiplets, and that these dominant irreps guide an efficient truncation of the
\textit{ab initio} model space.

Several directions follow naturally from these results. The most pressing is a systematic
strategy of basis truncation together with a controlled estimate of the truncation error, turning the
observed irrep dominance into a predictive and improvable approximation. A second direction
pairs the symmetry-adapted truncation with low-momentum interactions, where SRG-evolved
potentials embedded in the SAM basis preserve the accelerated convergence of the binding
energy~\cite{Dang2026} and thereby extend the reach of the method toward heavier systems. A
third direction concerns the weak processes; since the GT operator is itself an
SU(4) generator, the $\beta$-decay and electron-capture rates that govern $r$-process
nucleosynthesis
are naturally sensitive to the underlying supermultiplet structure and are now within first-principles reach.
A fourth direction extends the symmetry-informed machine learning beyond nuclear masses by
deploying the symmetry-inspired operators and interpretable architecture to other observables 
that are shaped by the same deformation and spin-isospin structure.

The centennial of Wigner's 1937 paper arrives in 2037. The trajectory from empirical
observation to QCD-grounded understanding and now to \textit{ab initio} implementation
appears to be converging rapidly, and the directions outlined here chart a course for the
coming decade.

\vspace{6pt}

\section*{Author Contributions}
Conceptualization, P.D.; Investigation, P.D. and F.P.; Writing -- Original Draft Preparation, P.D.;
Writing -- Review \& Editing, P.D. and F.P.; Visualization, P.D.

\section*{Funding}
PD was supported by the Institute of International Education through the Quad Fellowship, 
Louisiana State University, Quantum CodeX, and the National Science Foundation through the Open 
Questions and Research Tools in Nuclear Astrophysics summer school under Grant No. OISE-1927130 
(IReNA). FP was supported by the National Natural Science Foundation of China under Grant No. 12175097.

\section*{Data Availability}
No new data were created or analyzed in this study. Data sharing is not applicable to this article.

\begin{acknowledgments}
This review is dedicated to the memory of Dr. Jerry P. Draayer, who passed away
on September 13th, 2026 and whose seminal work on Elliott's $\rm SU(3)$ model, Symplectic $\rm Sp(3,R)$ 
model, SA-NCSM, and most recently, SAM has and will continue to influence the state of the art of
nuclear structure theory. The authors acknowledge Daniel Langr and Tomas Dytrych for \texttt{DeepSANE}'s
code development and are grateful to Serdar Elhatisari and Dean Lee for providing lattice phase shifts. 
Discussions with Anna McCoy on transition selection rules, Catherine Deibel on $\beta$-decay data, 
and Martin Savage and David Kaplan on the entanglement aspect are all appreciated.
\end{acknowledgments}

\section*{Conflicts of Interest}
The authors declare no conflicts of interest.

\section*{Abbreviations}

The following abbreviations are used in this manuscript:\\

\noindent
\begin{tabular}{@{}ll}
3DHO & Three-Dimensional Harmonic Oscillator \\
CC & Coupled Cluster \\
EFT & Effective Field Theory \\
$\chi$EFT & Chiral Effective Field Theory \\
GT & Gamow--Teller \\
IAS(s) & Isobaric Analog State(s) \\
IBM & Interacting Boson Model \\
IMSRG & In-Medium Similarity Renormalization Group \\
irrep(s) & Irreducible Representation(s) \\
MUSY & Multiconfigurational Dynamical Symmetry \\
NCSM & No-Core Shell Model \\
NLEFT & Nuclear Lattice Effective Field Theory \\
nn & neutron-neutron \\
np & neutron-proton \\
pp & proton-proton \\
QCD & Quantum Chromodynamics \\
QMC & Quantum Monte Carlo \\
RME(s) & Reduced Matrix Element(s) \\
SA-NCSM & Symmetry-Adapted No-Core Shell Model \\
SAM & Symmetry Adapted Model \\
SM & Standard Model \\
SpNCCI & Symplectic No-Core Configuration Interaction \\
SRG & Similarity Renormalization Group
\end{tabular}


\bibliography{References}

\end{document}